\documentclass[final,3p,times]{elsarticle}

\usepackage{amssymb}
\usepackage{amsmath}
\usepackage{amsthm}
\usepackage{amsfonts}
\usepackage{mathtools}
\usepackage{dsfont}
\usepackage{bm}
\usepackage{graphicx}
\usepackage{subcaption}
\usepackage{tikz}
\usetikzlibrary{calc, positioning, fit}
\usepackage{cleveref}
\crefformat{equation}{(#2#1#3)}
\usepackage{xcolor}

\usepackage[ruled,vlined,linesnumbered]{algorithm2e}

\newcommand{\flSet}{\mathcal{F}}

\newcommand{\mnorm}[1]{{\left\vert\kern-0.25ex\left\vert\kern-0.25ex\left\vert #1 
    \right\vert\kern-0.25ex\right\vert\kern-0.25ex\right\vert}}

\newtheorem{definition}{Definition} 
\newtheorem{theorem}{Theorem}
\newtheorem{lemma}{Lemma}

\newcommand{\ie}{{\it i.e.}}

\newcommand{\fix}[1]{{#1}}

\newcommand{\intSet}{\mathbb{Z}}
\newcommand{\realSet}{\mathbb{R}}

\newcommand{\actSet}{\mathcal{A}}

\newcommand{\plSet}{\mathbf{M}}
\newcommand{\xVec}{\bm{x}}
\newcommand{\BR}{\text{BR}}
\usepackage{lineno}
\usepackage{booktabs}

\journal{Transportation Research Part C}

\begin{document}
\begin{frontmatter}

\title{Game-theoretic Regulated Decentralized Coordination for \\ Airspace Sector Overload Mitigation}

\author[utexas]{Jaehan Im} 
\author[enac]{Daniel Delahaye}
\author[utexas]{David Fridovich-Keil}
\author[utexas]{Ufuk Topcu}

\affiliation[utexas]{%
    organization={The University of Texas at Austin},
    addressline={110 Inner Campus Drive},
    city={Austin},
    postcode={78712},
    state={Texas},
    country={United States}
}

\affiliation[enac]{
    organization = {École Nationale de l’Aviation Civile},
    addressline = {7 Avenue Édouard Belin, CS 54005},
    city = {Toulouse Cedex 4},
    postcode = {31055},
    country = {France}
}

\begin{abstract}
Decentralized air traffic management systems offer a scalable alternative to centralized control, but often assume high levels of cooperation. 
In practice, such assumptions frequently break down since \fix{agents, each representing an airspace sector,} may operate independently and prioritize local objectives.
We address the problem of sector overload in decentralized air traffic management by proposing a regulated decentralized protocol that models self-interested behaviors based on best response dynamics. 
\fix{Each agent adjusts the departure times of its assigned flights} to reduce its own congestion, without requiring centralized joint optimization. 
A tunable cooperativeness factor models the degree to which each \fix{agent} accounts for overload in other sectors, while a minimal admissibility rule prevents local updates from creating new overloads.
\fix{We show that this regulatory restriction is sufficient to guarantee finite-time convergence of best response dynamics to a pure Nash equilibrium under partial cooperation.} 
In addition, we identify a sufficient condition under which an overload-free solution corresponds to a global minimizer of the potential function. 
Numerical experiments using \fix{one month of European flight data} demonstrate that the proposed algorithm substantially reduces overload even with only minimal cooperation, \fix{achieving performance close to a centralized global mixed-integer optimizer and exceeding a centralized first-come-first-served baseline while requiring substantially less computation than centralized optimization. 
We further evaluate departure delay and fairness to quantify the operational cost of overload mitigation.}
\end{abstract}




\end{frontmatter}


\section{Introduction} \label{section:Introduction}

Decentralized approaches to air traffic management (ATM) are gaining attention as \fix{air traffic systems become more complex} \cite{GameTheoretic_ATM, Leveraging_ATM, DATFM_Perf_and_Fault, DATFM_Multi_agent_control}, \fix{alongside the emergence of advanced air mobility (AAM) operations involving numerous independent or autonomous agents} \cite{DATFM_for_AAM, FAA_UAM_ConOps}.
These approaches often rely on \emph{collaborative decision-making}, which assumes that stakeholders are willing to share information and cooperate toward a joint solution \cite{GameTheoretic_ATM,CDM_conf_paper,CDM_Cooperation_Design,CDM_Evolutionary,CDM_reroute}. 
\fix{However, the degree of cooperation that can be assumed may be limited when stakeholders have different or conflicting objectives, particularly across organizational or regional boundaries \cite{ADIZ_conflict,CoopFail_sector_conflict,FIR_agreement_conflict,FIR_departure_conflict}. 
This motivates studying what regulatory structure is necessary to maintain desirable system-level behavior while preserving distributed decision authority among self-interested agents.}

A key operational challenge in such decentralized settings is managing sector overload, which occurs when the number of aircraft in a sector exceeds a given capacity \cite{overload_complexity,sector_overload_workload,overload_capacity}. 
This leads to safety risks and increased controller workload \cite{sector_overload_workload,overload_capacity}, and typically requires flight plan adjustments \cite{CDM_reroute,rerouting_adacher,Rerouting}.
These interdependencies create a multi-agent decision-making problem, where each sector’s action influences the congestion experienced by others.
\fix{We consider a conceptual model of distributed coordination in future ATM operations. 
The airspace configuration and sector capacities are treated as given, while decision authority over traffic schedule adjustments is distributed among interacting agents.}

We propose a regulated decentralized mechanism that mitigates sector overloads by modeling sector interactions through a game-theoretic framework, capturing the full range from self-interested to cooperative behaviors. 
\fix{Each agent updates its decisions using best-response dynamics, myopically adjusting the departure times of its assigned flights to reduce its own congestion.} 
A minimal admissibility rule prevents local updates from introducing new overloads. 
To capture varying levels of cooperative behavior, we introduce an adjustable \textit{cooperativeness factor}~$\kappa$, which interpolates between fully self-interested ($\kappa = 0$) and fully cooperative ($\kappa = 1$) behavior. 
This formulation enables us to analyze how limited cooperation \fix{and regulatory intervention} affect overall system performance without requiring centralized joint optimization.

We provide theoretical guarantees on the convergence of best-response dynamics in this setting. Specifically, we show that for $\kappa = 1$, the game is a potential game, and best-response dynamics converges to a pure Nash equilibrium unconditionally. For intermediate values ($0\leq\kappa<1$), we prove that best-response dynamics converge in finitely many rounds under a mild restriction on update rules: \fix{agents} must avoid actions that introduce new overloads in other sectors. Additionally, we identify a sufficient condition under which any feasible solution (one with no sector overload) corresponds to a global minimizer of the potential function, providing insight into when the algorithm yields overload-free outcomes.

We validate the approach through numerical experiments using \fix{one month} of real flight data from the \fix{BREST flight information region in European airspace}. 
\fix{The observed traffic and sector geometry are used as a realistic test environment for the proposed coordination mechanism.} 
Results show that even with a minimal level of \textit{self-prioritizing} cooperation---where each \fix{agent} agrees to a cooperative decision only if it does not compromise its own overload---the proposed decentralized algorithm \fix{substantially reduces sector overload and approaches the centralized solver while preserving distributed decision authority}. 
\fix{The proposed algorithm also requires substantially less computation than centralized mixed-integer optimization, particularly under tighter capacity constraints. We further report departure delay and fairness metrics to quantify the operational cost of overload mitigation.}

In summary, the proposed algorithm \fix{addresses sector overload mitigation in an intermediate regime between unrestricted decentralized adaptation and centralized system-wide optimization. We characterize how self-interested best-response dynamics interact with limited regulatory rules, establish convergence and overload-related guarantees, and demonstrate using real-world data that significant overload reduction can be achieved even with only very limited cooperation.} The remainder of the paper reviews related work, presents the model and theoretical results, describes the experimental setup, and reports the numerical findings before concluding.

\section{Related Work}

\subsection{Centralized approaches in ATM}

A core challenge in ATM is \emph{demand--capacity balancing}, the process of aligning air traffic demand with sector capacities to prevent overload \fix{\cite{Leveraging_ATM}}. 
Centralized approaches rely on a central authority with access to system-wide information to prevent congestion and coordinate schedules across sectors \fix{\cite{central_optimization_atfm,bertsimas1998air}}. 
Prominent methods include air traffic flow management \fix{\cite{bertsimas1998air,central_optimization_atfm,Rerouting}} and performance-based ATM \fix{\cite{GameTheoretic_ATM}}, which often use optimization models to minimize delay and promote fairness \cite{Leveraging_ATM}. 
While effective in controlled settings, centralized schemes face scalability challenges as traffic grows and operations become more complex and heterogeneous, and such centralized ATFM is not feasible in some regions---Asian airspace being a representative example \fix{\cite{FIR_agreement_conflict,FIR_departure_conflict}}.

\subsection{Decentralized ATM and collaborative frameworks}

To address these limitations, decentralized concepts have been explored through \fix{decentralized control architectures \cite{DATFM_Perf_and_Fault} and multi-agent coordination \cite{DATFM_Multi_agent_control}}. 
\emph{Collaborative decision-making} is a representative approach in this area, promoting \fix{joint decision processes among ATM stakeholders \cite{CDM_conf_paper} and cooperative air--ground procedures \cite{CDM_Cooperation_Design}} to improve outcomes---indeed collaborative decision-making has been successfully deployed in the United States \cite{CTOP} and the \emph{Airport-CDM} systems at major European airports \cite{A_CDM}, supported by the network manager in Brussels.
However, many decentralized approaches assume a high degree of transparency and willingness to cooperate---assumptions that often fail in practice due to competitive dynamics and the noncooperative nature of ATM stakeholders, particularly in international or boundary airspace sectors \fix{\cite{CoopFail_sector_conflict, FIR_agreement_conflict}}.

\subsection{\fix{Noncooperative and regulated coordination}}

\fix{Game-theoretic approaches have been used to model airline competition and congestion \cite{noncoop_hansen, freq_competition_airport_congestion}, performance-based ATM \cite{GameTheoretic_ATM}, traffic forecasting \cite{game_forecasting}, and multi-agent traffic management \cite{gameTheo_atm}. Related studies have also applied game-theoretic concepts to air traffic control training \cite{potential} and network-wide safety--efficiency optimization \cite{Optimizing_nationwide}. Prior work considers several distinct forms of noncooperative coordination. In \cite{DTM_NE}, tactical free-flow traffic management is formulated as a Nash game in which aircraft independently select trajectories. In \cite{RRCE}, reduced-rank correlated equilibria provide a recommendation-based coordination mechanism for general multiplayer matrix games. In \cite{TACo}, self-interested agents select an outcome through an explicit trading-based negotiation mechanism. These approaches address different coordination settings and do not study regulated best-response dynamics for sector overload mitigation. 
A more closely related study \cite{im2027iterative} combines trading-based negotiation with taxation-like regulatory oversight, where taxes, trading assets, and iterative coordination signals reshape agents' negotiated outcomes. In contrast, the present work studies protocol-level oversight that directly restricts admissible unilateral best-response updates.}

\subsection{\fix{Demand-side and capacity-side overload mitigation}}

\fix{Sector overload may be addressed through either demand-side or capacity-side interventions. Demand-side approaches modify traffic flows through departure delay, rerouting, or related traffic-flow measures \cite{Leveraging_ATM, CDM_reroute, Rerouting}, while dynamic airspace configuration and sector opening, collapsing, or redesign modify the available capacity itself \cite{romero2025smarts, tang2012multi}. The present work focuses on demand-side coordination by treating the airspace configuration and sector capacities as given, thereby isolating the interaction between self-interested best-response behavior and regulatory intervention. Capacity-side reconfiguration may be combined with the proposed framework in future work.}


\section{Airspace Sector Overload Problem}
\label{sec:problem}

\begin{figure}[bp!]
    \centering
    \includegraphics[width=0.9\linewidth]{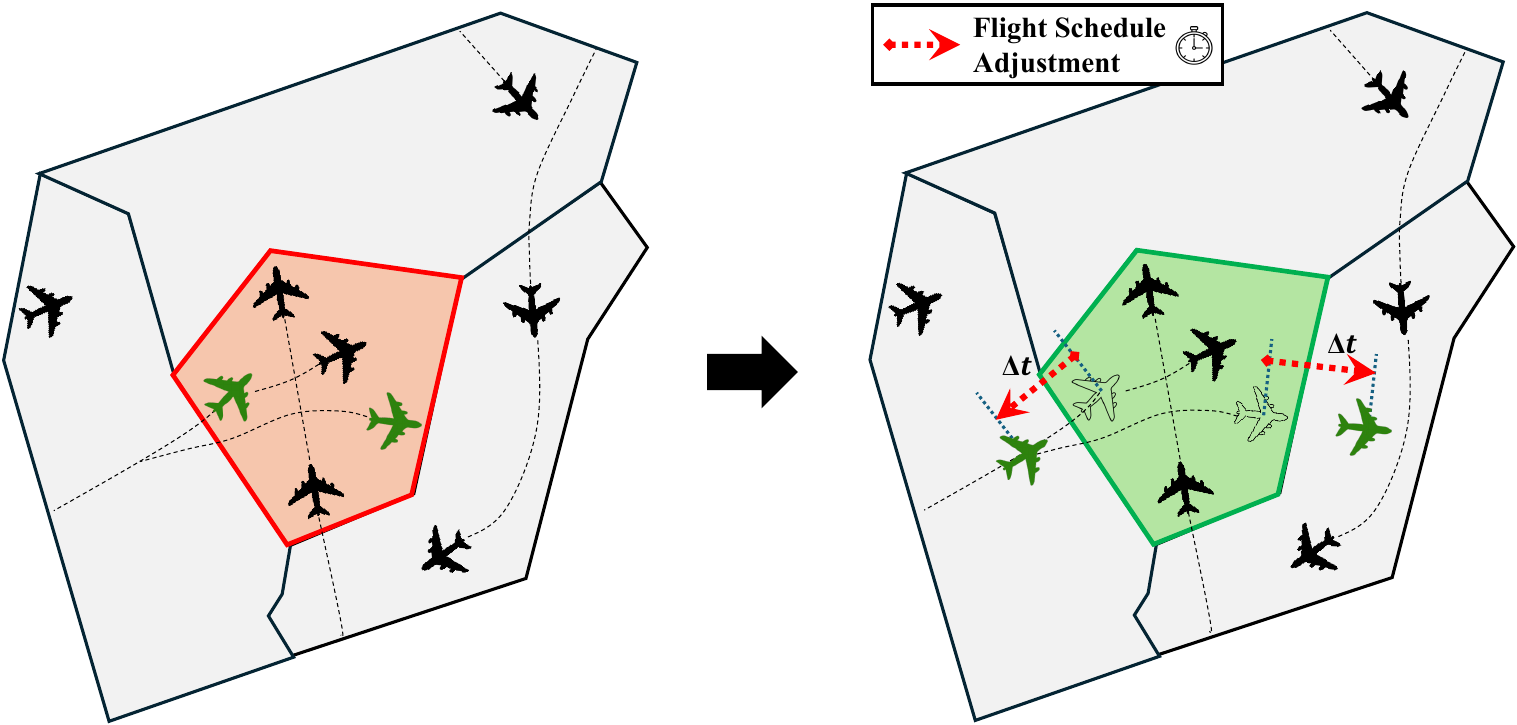}
    \caption{
        Illustration of sector overload mitigation. 
        (Left) A sector with a capacity limit of three aircraft is overloaded with five active flights (red region). 
        (Right) Two flights (green) adjust their flight schedule (red dashed arrows), 
        removing the overload and restoring the sector to a feasible state (green region).
    }
    \label{fig:concept}
\end{figure}

We formulate the sector overload resolution problem in terms of the agents, their decision variables, and the capacity constraints that should be respected. \Cref{fig:concept} illustrates a simple example: a sector exceeding its capacity and the flight schedule adjustments removing overloads.

\subsection{Airspace system and agents}
The airspace is partitioned into $m$ sectors \fix{represented by $m$ decision-making agents.} We denote the set of agents by $\plSet = \{1,\ldots,m\}$. \fix{Each flight is assigned to exactly one agent for scheduling, and $\mathcal{F}_i$ denotes the set of flights assigned to agent $i$. Thus, the sets $\{\mathcal{F}_i\}_{i\in\plSet}$ form a partition of the flight set, with $n_i = |\mathcal{F}_i|$ and $n = \sum_{i\in\plSet} n_i$. The assignment of flights to agents is scenario-dependent. 
For example, an agent may schedule flights associated with a particular departure region or entering through a designated sector.}
Each \fix{agent} can adjust the departure times of its flights within a finite action set 
$\actSet \subset \realSet$, consisting of $p$ discrete scheduling options. 
For example, $\actSet = \{0, 5, 10, 15, 20, 25, 30\}$ (minutes) represents departure-time adjustments 
in 5-minute increments over a 0 to 30 minute range. 
The decision variable of \fix{agent} $i$ is then $x_i \in \actSet_i := \actSet^{n_i}$, 
and the joint action profile across all \fix{agents} is $\xVec = (x_1,\ldots,x_m) \in \prod_{i\in\plSet}\actSet_i = \actSet^{\sum_i n_i} = \actSet^{n}$.

\subsection{Occupancy metric and overload definition}
We measure sector overload using the \emph{occupancy metric} defined as the number of aircraft present in a sector at a given time \cite{occupancy_count_1, occupancy_count_2}. Let $\mathcal{T}$ denote the set of discrete time bins in the planning horizon, with bin length $\Delta \tau$. Let $D_i \in \intSet_{>0}$ denote the capacity of sector $i$, and let $C_{ji}^{\tau} : \actSet^n \to \realSet_{\ge 0}$ represent the contribution of flights scheduled by \fix{agent} $j$ to the occupancy of sector $i$ at time bin $\tau$.

More specifically, if $\mathcal{F}_j$ denotes the set of flights scheduled by \fix{agent} $j$, then
\begin{equation}
    C_{ji}^{\tau}(\xVec)
    =
    \sum_{f\in\mathcal{F}_j}
    \mathbf{1}\left\{
    \tau \in [e_{fi}(x_f), q_{fi}(x_f))
    \right\},
\end{equation}
\fix{where $\mathbf{1}\{\cdot\}$ is the indicator function}, and $e_{fi}(x_f)$ and $q_{fi}(x_f)$ are the entry and exit times of flight $f$ in sector $i$ under departure-time adjustment $x_f$, respectively. \fix{Specifically, $e_{fi}(x_f)=e_{fi}(0)+x_f$ and $q_{fi}(x_f)=q_{fi}(0)+x_f$ and sector membership is defined with respect to the full three-dimensional sector geometry.}
If flight $f$ does not traverse sector $i$, the corresponding indicator is zero. Thus, dwell time is incorporated because a flight contributes to the occupancy of sector $i$ for every time bin during which it remains inside the sector.

The total occupancy of sector $i$ at time bin $\tau$ is $\sum_{j\in\plSet} C_{ji}^{\tau}(\xVec)$. The overload of sector $i$ at time bin $\tau$ is
\begin{equation}
    \ell_i^\tau(\xVec)
    =
    \max\left(0,\; \sum_{j\in\plSet} C_{ji}^{\tau}(\xVec) - D_i\right).
\end{equation}
The \emph{cumulative overload} in sector $i$, $L_i : \actSet^n \to \realSet_{\ge 0}$, is then defined as:
\begin{equation}\label{eq:overload}
    L_i(\xVec)
    =
    \sum_{\tau\in\mathcal{T}}
    \ell_i^\tau(\xVec)\Delta \tau,
    \quad \xVec \in \actSet^n.
\end{equation}

\noindent
A solution $\xVec^\star \in \actSet^n$ is \emph{feasible} if all sectors respect their capacity constraints at every time bin, i.e.,
\begin{equation}\label{eq:constraint}
    \sum_{i\in\plSet} L_i(\xVec^\star) = 0.
\end{equation}

\noindent
In other words, \Cref{eq:overload} measures a specific sector's cumulative overload over the planning horizon, and \Cref{eq:constraint} defines feasibility as complete elimination of overload across all sector-time pairs.

\subsection{Action updates and local effects}
Consider a unilateral update by agent $i$, denoted $\xVec_i' = (x_1,\ldots,x_i',\ldots,x_m)$ for $x_i' \in \actSet_i$. 
Under the occupancy metric, the update affects only the occupancy contributions generated by flights scheduled by \fix{agent} $i$. 
Accordingly, for every time bin $\tau \in \mathcal{T}$,
\begin{equation}
    C_{jk}^{\tau}(\xVec) = C_{jk}^{\tau}(\xVec_i'), 
    \quad \xVec \in \actSet^n, \quad \forall j\ne i,\; \forall k\in\plSet .
    \label{eq:Cjk_invariance}
\end{equation}
Thus, only $C_{ik}^{\tau}$ for $k\in\plSet$ may change when agent $i$ updates its decision.
This property will be central to the game-theoretic analysis in the next section.

\subsection{Operational requirements and assumptions} \label{sec:requirements}
The mechanism we seek must align with operational realities of regulated decentralized ATM:
\begin{itemize} \setlength{\itemindent}{-0.8em}
    \item \textbf{Local decision-making:} Each \fix{agent} independently chooses schedule adjustments for flights under its control. We assume that flight-plan information is shared so that sector-time occupancies and overloads can be computed. \fix{The mechanism also assumes a common sequential update order and system-wide overload evaluation.}
    \item \fix{\textbf{Pure-strategy outcome:}} The resulting decisions must form a pure Nash equilibrium, ensuring predictable and stable schedules as required in ATM operations. Mixed-strategy equilibria would yield randomized schedules, which are not acceptable in safety-critical systems.
    \item \fix{\textbf{Finite-time convergence:}} The mechanism must guarantee convergence within a finite number of steps to allow timely coordination. \fix{Finite-time convergence does not by itself imply that the resulting equilibrium is overload-free, and the conditions for overload-free outcomes are treated separately in the theoretical analysis.}
\end{itemize}

\subsection{Problem definition}

Based on this structure, we define the \emph{sector overload problem} as determining a joint action profile $\xVec^\star$ that eliminates all overload, i.e., identify departure-time adjustments for all flights that satisfies the feasibility condition in \Cref{eq:constraint}. Building on this formulation, the next section introduces our game model and establishes its theoretical properties.

\section{Game-Theoretic Model of Sector Interaction} \label{sec:proof}

We model the sector overload problem as a game in which each \fix{decision-making entity is treated as an individual agent}. We adopt \emph{best-response dynamics} as the update rule: each \fix{agent} independently adjusts its own schedule using shared flight-plan information to evaluate occupancy and overload without relying on centralized joint optimization. \fix{Best-response dynamics provides a model of reactive local adaptation in the distributed coordination setting considered here.}

\begin{definition}[Best-response dynamics \cite{book_shoham,swensonBRD}] \label{def:brd}
For a game involving $m$ agents with cost functions 
$J_i : \actSet^n \to \realSet$ for $i \in \{1,\ldots,m\}$ 
and action sets $\{\actSet_i\}_{i=1}^m$, 
the best-response mapping of agent~$i$, $\BR_i : \actSet^n \to \actSet_i$, 
is defined as:
\begin{equation}\label{eq:br_mapping}
    \BR_i(\xVec) := \arg\min_{a \in \actSet_i} J_i(a, \xVec_{-i}), 
    \quad \xVec \in \actSet^n,
\end{equation}
where $\xVec_{-i} = (x_1,\ldots,x_{i-1},x_{i+1},\ldots,x_m)$.
The best-response dynamics generates a sequence $\{\xVec^{(t)}\}_{t \ge 0}$ such that, at round $t$, one agent~$i$ unilaterally updates its decision according to
\begin{equation}\label{eq:br_update}
    x_i^{(t+1)} \in \BR_i(\xVec^{(t)}).
\end{equation}
\end{definition}
\noindent
The best-response dynamics is known to converge to a pure Nash equilibrium in \emph{potential games} \cite{potential_opg, book_shoham, swensonBRD}, provided that the action space is finite, as in our case. 

\begin{definition}[Potential game \cite{book_shoham}]
A game involving $m$ agents with cost functions $\{J_i : \actSet^n \to \realSet\}_{i=1}^m$ is a \emph{potential game} if there exists a \emph{potential function} $\Phi : \actSet^n \to \realSet$ such that, for any agent $i$ and any $x_i' \in \actSet_i$, where $\xVec_i' = (x_1,\ldots,x_i',\ldots,x_m)$,
\begin{equation}
    J_i(\xVec_{i}') - J_i(\xVec) \;=\; \Phi(\xVec_i') - \Phi(\xVec), \quad \xVec \in \actSet^n.
\end{equation}
That is, the change in an agent’s cost from a unilateral deviation exactly matches the change in the potential function. The potential function provides a global measure aligned with individual incentives, so unilateral improvements correspond to decreasing this function. Hence, its local minima coincide with pure Nash equilibria.
\end{definition}
\noindent
Since pure Nash equilibria correspond to reliable, deterministic decisions, showing that our model satisfies a potential function will ensure that it meets the operational requirements of ATM systems.

The remainder of this section develops as follows: 
(i) We first introduce a cost function with a tunable cooperativeness factor $\kappa$. 
(ii) We then establish conditions under which the game admits a potential structure and prove finite termination under a mild restriction for $0\leq\kappa<1$.
(iii) Finally, we prove that best-response dynamics converges to a pure Nash equilibrium and identify when overload-free solutions are global minimizers of the potential function.

\subsection{Cost function and cooperativeness factor $\kappa$}
We define the cost function of sector $i \in \plSet$, $J_i : \actSet^n \to \realSet$, as
\begin{equation}\label{eq:cost}
    J_i(\xVec) \;=\; L_i(\xVec) \;+\; \kappa \sum_{j \in \plSet \setminus \{i\}} L_j(\xVec), \quad \xVec \in \actSet^n,
\end{equation}
where $L_i(\xVec)$ is the overload in sector $i$ defined in \Cref{eq:overload}, and $0 \le \kappa \le 1$ is the cooperativeness factor. 
The first term reflects each agent’s self-interest, while the second introduces a penalty for system-wide overload.
This formulation interpolates between fully self-interested behavior ($\kappa=0$) and fully cooperative behavior ($\kappa=1$).
In particular, we identify a special regime called 
\emph{self-prioritizing cooperativeness}. 

\begin{definition}[Self-prioritizing cooperative behavior]
An agent $i$ exhibits self-prioritizing cooperative behavior 
if it adopts a cooperative action only when that action does not increase 
its own overload; that is,
\begin{equation}
   L_i(\xVec_i'(a)) \le L_i(\xVec), \quad \xVec \in \actSet^n,\,\forall a \in \actSet_i,
\end{equation}
where $\xVec_i'(a) = (x_1, \ldots ,x_{i-1},a,x_{i+1},\ldots,x_m)$.
\end{definition}
The following theorem provides a sufficient condition on $\kappa$ under which an agent exhibits this behavior.

\begin{theorem}[Sufficient bound on $\kappa$ for self-prioritization]\label{thm:self-prio}
\fix{For each agent $i\in\plSet$, define the total dwell time of its assigned flights in sectors other than $i$ as
\begin{equation}\label{eq:Wi}
    W_i :=
    \sum_{f\in\mathcal{F}_i}
    \sum_{j\in\mathcal{S}_f\setminus\{i\}}
    \bigl(q_{fj}(0)-e_{fj}(0)\bigr),
\end{equation}
where $\mathcal{S}_f$ denotes the set of sectors traversed by flight $f$. If
\begin{equation}\label{eq:kappa-bound}
    \kappa < \frac{\Delta\tau}{\max_{i\in\plSet} W_i},
\end{equation}
where $\Delta\tau$ is the time-bin length, then no agent will adopt an action that increases its own cumulative overload.}
\end{theorem}

\noindent
This condition ensures that the cooperative term in \Cref{eq:cost} never outweighs the self-interest term, and therefore agents remain self-prioritizing.
The bound remains conservative because \fix{it assumes that the entire dwell time of the flights assigned to an agent in all other sectors could contribute to reducing overload.}
The self-prioritizing regime represents a minimal form of cooperation beyond purely self-interested behavior: \fix{agents} are willing to take cooperative actions only when such actions do not worsen their own overload. 

\subsection{Potential game properties}
We now examine whether the game defined by \Cref{eq:cost} admits a potential function. 
We begin by considering the case where the set of overloaded sector-time pairs remains fixed.

\begin{theorem}[Potential game under fixed overloaded sector-time set]\label{thm:potential-fixed}
Let $\mathcal{O} : \actSet^n \to 2^{\plSet \times \mathcal{T}}$ denote the \emph{set of overloaded sector-time pairs}, defined by:
\begin{equation}
    \mathcal{O}(\xVec)
    :=
    \{\, (i,\tau) \in \plSet \times \mathcal{T}
    \mid \ell_i^\tau(\xVec) > 0 \,\},
    \quad \xVec \in \actSet^n .
\end{equation}
If this set of overloaded sector-time pairs remains unchanged under unilateral deviations, then the game with cost function \Cref{eq:cost} admits the potential function $\Phi: \actSet^n \to \realSet$,
\begin{equation} \label{eq:potential}
    \Phi(\xVec)
    =
    \kappa \sum_{i \in \plSet} L_i(\xVec) 
    +
    (1-\kappa)
    \sum_{(i,\tau)\in \mathcal{O}(\xVec)}
    C_{ii}^{\tau}(\xVec)\Delta\tau,
    \quad \xVec \in \actSet^n .
\end{equation}
\begin{proof}
    The proof is provided in~\ref{app1}.
\end{proof}
\end{theorem}

\noindent
This means that, as long as unilateral deviations do not alter which sector-time pairs are overloaded, 
each agent's cost improvement aligns with a decrease in the global potential function.

\begin{lemma}[Potential game for $\kappa=1$]\label{lem:kappa-one}
When $\kappa=1$, the game defined by \eqref{eq:cost} admits a potential function unconditionally.
\begin{proof}
    The proof is provided in \ref{app1}.
\end{proof}
\end{lemma}

\noindent
Unlike \Cref{thm:potential-fixed}, this result holds regardless of how the overloaded sector-time set changes. 
In particular, for $\kappa=1$ the potential function reduces to measuring only the system-wide cumulative overload, \ie\, 
$\Phi(\xVec)=\sum_{i\in\plSet}L_i(\xVec) \ge 0$, 
so every feasible solution $\xVec^\star$ satisfying \Cref{eq:constraint} is a global minimizer of $\Phi$.

\subsection{Termination guarantee and restricted mechanism}

When $0 \le \kappa < 1$, the game in \Cref{eq:cost} is not generally an unconditional potential game. 
However, we can guarantee finite termination of the restricted best-response dynamics by imposing a sufficient admissibility restriction on agents' actions.

For a joint action profile $\xVec$ and a unilateral deviation $\xVec_i'$, 
let $\mathcal{O}_f : \actSet^n \times \actSet^n \to 2^{\plSet \times \mathcal{T}}$ denote 
the set of sector-time pairs that become overloaded after the deviation:
\begin{equation}\label{eq:feasible_set_mapping}
    \mathcal{O}_f(\xVec,\xVec_i') 
    :=
    \{\, (j,\tau) \in \plSet\times\mathcal{T}
    \mid \ell_j^\tau(\xVec) = 0,\; \ell_j^\tau(\xVec_i') > 0 \,\}, 
    \quad \xVec \in \actSet^n,
\end{equation}
where $\xVec_i' := (x_1,\ldots,x_i',\ldots,x_m)$ for some $x_i' \in \actSet_i$. We impose the following restriction.

\begin{definition}[No-new-overload restriction] \label{def:noNewOvld}
Agents are restricted to actions $\xVec_i'$ such that $\mathcal{O}_f(\xVec,\xVec_i') = \emptyset$, \ie, no previously feasible sector-time pair becomes overloaded.
\end{definition}

\noindent
Under this restriction, a unilateral deviation can only maintain or reduce the set of overloaded sector-time pairs but can never expand it. \fix{As assumed in \Cref{sec:requirements}, agents have access to the system-wide overload information needed to evaluate this condition.} This allows us to establish the following property.

\begin{lemma}[Monotonic decrease of the overloaded sector-time set]\label{lem:monotone}
Under the no-new-overload restriction in \Cref{def:noNewOvld}, if a unilateral deviation changes the overloaded sector-time set, that is, 
$\mathcal{O}(\xVec_i') \neq \mathcal{O}(\xVec)$, 
then the set must strictly decrease:
\begin{equation}
    \mathcal{O}(\xVec_i') \subsetneq \mathcal{O}(\xVec), 
    \quad \xVec \in \actSet^n.
\end{equation}
\begin{proof}
The proof is provided in \ref{app1}.
\end{proof}
\end{lemma}

\noindent
\Cref{lem:monotone} implies that $\mathcal{O}(\xVec)$ either remains unchanged or strictly decreases after each round. 
Since $\plSet\times\mathcal{T}$ is finite, this monotone property directly implies that the overloaded sector-time set can change only finitely many times.

\begin{theorem}[Finite termination for $0\le\kappa<1$] \label{thm:termination}
Consider the game defined by the cost function~\eqref{eq:cost} with $0\le\kappa<1$. 
Under the no-new-overload restriction (\Cref{def:noNewOvld}), 
the restricted best-response dynamics terminates in finite time and converges to a pure Nash equilibrium with respect to the admissible action set induced by \Cref{def:noNewOvld}.
\begin{proof}
The proof is provided in \ref{app1}.
\end{proof}
\end{theorem}

\noindent
The termination follows from two observations. Let $\xVec^{(t)}$ denote the joint action profile at round $t$. While $\mathcal{O}(\xVec^{(t)})$ remains fixed, the problem is a potential game and the best-response dynamics converges by \Cref{thm:potential-fixed}. 
If $\mathcal{O}(\xVec^{(t)})$ changes, \Cref{lem:monotone} ensures that $\mathcal{O}(\xVec^{(t+1)})$ is a strict subset of $\mathcal{O}(\xVec^{(t)})$. Since $\plSet\times\mathcal{T}$ is finite, the overloaded sector-time set can change only finitely many times, and the number of best-response iterations at any fixed $\mathcal{O}(\xVec^{(t)})$ is bounded.
Therefore, the process must terminate in finite time.

\subsection{Convergence of best-response dynamics}

The convergence of the best-response dynamics follows from the potential structure 
and the monotonic evolution of the overloaded sector-time set~$\mathcal{O}(\xVec)$. 
\Cref{lem:kappa-one} guarantees that the game is an unconditional potential game when $\kappa=1$, while \Cref{thm:termination} establishes finite termination for $0\le\kappa<1$ under the no-new-overload restriction. 
We summarize these results in the following theorem.

\begin{theorem}[Convergence under best-response dynamics]\label{thm:convergence}
Consider the best-response dynamics defined in \Cref{def:brd}. 
Based on \Cref{lem:kappa-one}, \Cref{thm:potential-fixed}, and \Cref{thm:termination}, the following statements hold:
\begin{enumerate}
    \item When $\kappa=1$, the game admits an unconditional potential function, and best-response dynamics converges to a pure Nash equilibrium in finite time. 
    \item When $0\le\kappa<1$, if agents are restricted from actions that introduce new overloads according to \Cref{def:noNewOvld}, the restricted best-response dynamics terminates in finite time and converges to a pure Nash equilibrium with respect to the admissible action set induced by \Cref{def:noNewOvld}.
\end{enumerate}
In both cases, best-response dynamics is guaranteed to converge in a finite number of steps under the stated conditions.
\begin{proof}
The proof is provided in~\ref{app1}.
\end{proof}
\end{theorem}

\noindent
\Cref{thm:convergence} establishes finite-time convergence guarantees for the proposed best-response dynamics. 
For $\kappa=1$, convergence follows unconditionally from the potential-game structure. 
For $0\le\kappa<1$, convergence is guaranteed under the no-new-overload restriction, which prevents unilateral updates from creating overload in previously feasible sector-time pairs. 
This provides the key operational guarantee for the regulated decentralized protocol: interactions based on best responses terminate in finite time and produce deterministic schedules under the stated admissibility condition.
\fix{However, finite-time convergence does not by itself guarantee that the resulting equilibrium is overload-free, particularly for $0\leq\kappa<1$. Feasibility is a separate property, considered in the following subsection.}

\subsection{Sufficient condition for feasibility}
Beyond convergence, we examine how feasible overload-free solutions relate to the potential function. 
When $\kappa=1$, cost function \Cref{eq:cost} and potential function \Cref{eq:potential} become identical, which leads to the following result.

\begin{theorem}[Feasible solutions as global minimizers]\label{thm:feasible}
When $\kappa=1$, any feasible solution $\xVec^\star$ satisfying $L_i(\xVec^\star)=0$ for all $i$ is a global minimizer of \Cref{eq:potential}, the potential function.
\begin{proof}
    The proof is provided in \ref{app1}.
\end{proof}
\end{theorem}

\noindent
\Cref{thm:feasible} shows that under full cooperation, whenever a feasible solution exists, it coincides with a global minimizer of the potential function.

\medskip
To summarize, our theoretical results directly address the operational requirements identified in \Cref{sec:problem}. \fix{These properties establish the theoretical basis for the regulated distributed coordination setting considered in this work.}

\section{Decentralized Overload Mitigation Mechanism}
Building on the game-theoretic model in \Cref{sec:proof}, 
we propose a regulated decentralized algorithm based on best-response dynamics, \fix{where each decision-making agent represents an airspace sector}, to address the sector overload problem.
Each \fix{agent} computes its own schedule update, while system-level overload information is used to enforce the no-new-overload restriction. 
Thus, the proposed mechanism lies between fully centralized optimization and unconstrained decentralized decision-making.
\fix{The sequential update order and system-wide admissibility check constitute protocol-level coordination, but the schedule update itself is determined independently by each agent.}

\begin{algorithm}[hbt!]
\caption{Decentralized overload mitigation}
\label{alg:dom}
\KwIn{\fix{Agents} $\plSet$, time bins $\mathcal{T}$, action sets $\{\actSet_i\}_{i\in\plSet}$, capacities $\{D_i\}_{i\in\plSet}$, cooperativeness factor $\kappa\in[0,1]$, initial schedule $\xVec^{(0)}$}

$t \gets 0$\;
Compute $L(\xVec^{(t)})$ and $J_i(\xVec^{(t)})$ for all $i$\;
$\text{Improved} \gets \textbf{true}$\;
\While{$\text{Improved}$}{
    $\text{Improved} \gets \textbf{false}$\;
    \For{$i \in \plSet$}{
        Solve \Cref{eq:br-problem} and obtain candidate action $x^{\mathrm{cand}}$\;
        $\xVec^{\mathrm{cand}} \gets (x_1^{(t)},\ldots,x_{i-1}^{(t)},x^{\mathrm{cand}},x_{i+1}^{(t)},\ldots,x_m^{(t)})$\;
        \If{$J_i(\xVec^{\mathrm{cand}}) < J_i(\xVec^{(t)})$}{
            $\xVec^{(t+1)} \gets \xVec^{\mathrm{cand}}$\;
            Update $L(\xVec^{(t+1)})$ and $J_j(\xVec^{(t+1)})$ for all $j\in\plSet$\;
            $\text{Improved} \gets \textbf{true}$\;
            $t \gets t + 1$\;
        }
        \If{$\sum_{j\in\plSet} L_j(\xVec^{(t)}) = 0$}{
            \Return{$\xVec^{(t)}$}
        }
    }
}
\Return{$\xVec^{(t)}$}
\end{algorithm}

\subsection{Algorithm overview}
At round $t$, \fix{agents} update sequentially. When \fix{agent} $i$ is active, it searches for a unilateral update $a$ that minimizes its cost $J_i$ in \Cref{eq:cost}, 
while keeping the other \fix{agents'} actions fixed. 
The candidate joint action is denoted by $\xVec_i^{(t+1)}(a)=(x_1^{(t)},\ldots,x_{i-1}^{(t)},a,x_{i+1}^{(t)},\ldots,x_m^{(t)})$. 
\fix{The best response is obtained by solving the discrete optimization problem over $\actSet_i$ subject to the no-new-overload restriction in \Cref{def:noNewOvld}. Evaluating this restriction uses the shared sector-occupancy information assumed in \Cref{sec:problem}.}
This process repeats until no agent can improve.

\subsection{Best response action update}
Given the other \fix{agents'} actions at round $t$, \fix{agent} $i$ computes its best response by solving
\begin{equation} \label{eq:br-problem}
\begin{aligned}
x_i^{(t+1)} \;\in\; \arg\min_{a\in\actSet_i} \quad & 
    L_i\big(\xVec_i^{(t+1)}(a)\big) + \kappa \sum_{j\in\plSet\setminus\{i\}} L_j\big(\xVec_i^{(t+1)}(a)\big) \\
\text{s.t.} \quad & \ell_j^\tau(\xVec)=0 \;\;\Longrightarrow\;\; \ell_j^\tau\big(\xVec_i^{(t+1)}(a)\big)=0, 
\quad \forall j\in\plSet,\; \forall \tau\in\mathcal{T}.
\end{aligned}
\end{equation}
That is, \fix{agent} $i$ minimizes its cost while respecting the restriction that no new overload may be introduced. This update rule corresponds to the best-response dynamics analyzed in \Cref{thm:convergence}.

\fix{In the numerical implementation, we formulate this local best-response problem as a mixed-integer linear program and solve it using the optimization procedure described in the experimental setup.}

\subsection{Termination condition}
The algorithm terminates when either of two conditions is met: 
(i) a full round over all \fix{agents} yields no update, meaning no agent can further reduce its cost and we have reached a pure Nash equilibrium under \Cref{thm:convergence}; or 
(ii) a feasible overload-free solution is found mid-round, enabling early termination.

\section{Numerical Experiments}
We conduct numerical experiments to evaluate the performance of the proposed algorithm.
The experiments are designed to answer three main questions:
(i) how the cooperativeness factor $\kappa$ affects algorithm performance,
(ii) how the algorithm compares against representative baseline methods, and
(iii) whether the method scales to scenarios with large numbers of flights.
We first describe the experimental setup and baselines before presenting the results.

\subsection{Experiment setup and data description} \label{sec:exp-setup}
\begin{figure}[hbt!]
    \centering
    \begin{subfigure}{0.99\linewidth}
        \centering
        \includegraphics[width=0.99\linewidth,trim=0 0 0 0,clip]{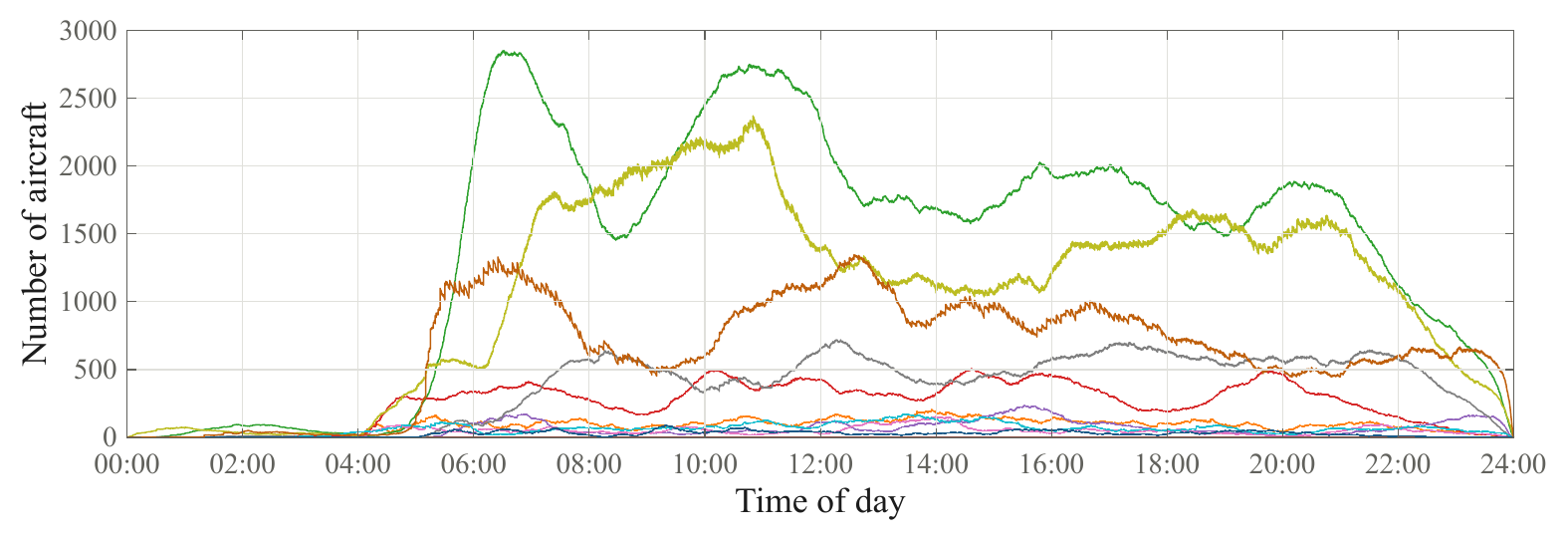}
        \caption{Traffic demand profile for the European airspace aggregated into 12 country-level groups. Each color represents one group.}
        \label{fig:TDP_Europe}
    \end{subfigure}
    \begin{subfigure}{0.99\linewidth}
        \centering
        \includegraphics[width=0.99\linewidth,trim=0 0 0 0,clip]{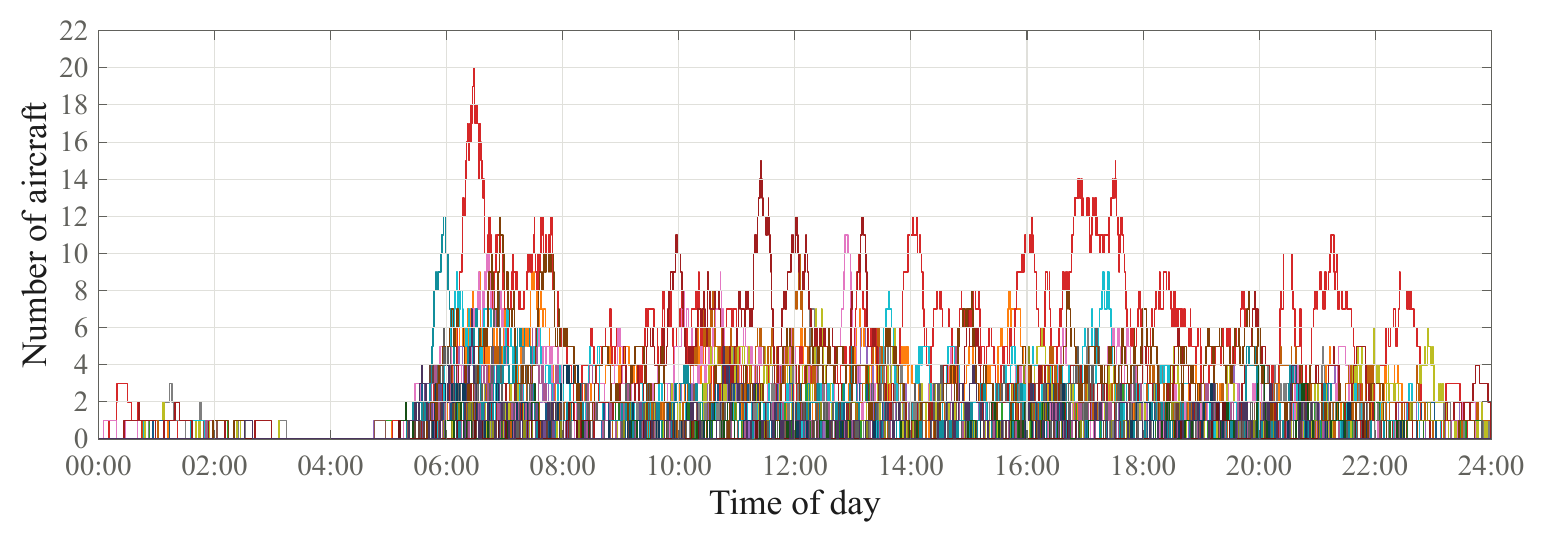}
        \caption{Traffic demand profile for the BREST FIR at the sector level. Each color represents one of the 28 sectors.}
        \label{fig:TDP_BREST}
    \end{subfigure}

    \caption{Traffic demand profiles on July 30, 2023.
    The plots illustrate temporal variations in traffic demand at two different levels of spatial aggregation:
    (a) European airspace grouped by 12 countries, and
    (b) BREST FIR divided into 28 individual sectors.}
    \label{fig:TDP_Whole}
\end{figure}

\fix{We evaluate the proposed algorithm using one month of flight data from European airspace.} Figure 2 illustrates July 30, 2023 as a representative day at two levels of aggregation, with the entire European airspace grouped into 12 country-level regions in Figure 2a and the BREST flight information region (FIR) in France divided into 28 sectors in Figure 2b. \fix{Our main experiments focus on the BREST FIR, using the traffic observed on each day of the one-month dataset.} Each sector is assigned a uniform capacity limit $D_i = 10$ for all $i\in\{1,\ldots,28\}$, and the occupancy is evaluated with one-second time bins (\ie, $\Delta\tau=1$ s).

Each flight’s action set is defined as discrete departure delays in 5-minute increments, ranging from $0$ to $30$ minutes: $\actSet = \{0,5,10,15,20,25,30\}$.
\fix{Following the formulation in \Cref{sec:problem}, each flight is assigned to the agent associated with the first BREST sector traversed by that flight. The entry and exit times $e_{fi}(0)$ and $q_{fi}(0)$ are obtained directly from the traffic data as the first and last times at which flight $f$ is recorded in sector $i$, respectively, where we recall that $e_{fi}(0)$ and $q_{fi}(0)$ denote the nominal entry and exit times of flight $f$ in sector $i$. Departure-time adjustments shift these times while preserving the observed sector sequence.}
\fix{Agent} $i$’s action is $x_i \in \actSet_i := \actSet^{n_i}$ where we recall that $n_i$ is the number of flights \fix{assigned to agent $i$}, and the joint action $\xVec$ specifies the departure schedule for all flights.
\fix{This assignment provides a conceptual distributed decision structure for evaluating the proposed framework.}

The decentralized algorithm is tested with cooperativeness factors $\kappa \in \{0,10^{-6},0.5,1\}$, covering regimes from fully self-interested to fully cooperative.
In particular, $\kappa=10^{-6}$ represents a small-cooperation setting \fix{within} the self-prioritizing regime in \Cref{thm:self-prio}. 
\fix{For the BREST dataset, the most restrictive bound across the evaluated days is $\kappa<2.34\times10^{-6}$, so $\kappa=10^{-6}$ satisfies the sufficient condition throughout the experiments.}
We include $\kappa=0.5$ as an intermediate value to illustrate trade-offs between this regime and the fully cooperative case.
\fix{The one-month dataset is used to evaluate performance across different daily traffic conditions, and the daily outcomes and their mean are reported in the corresponding figures.}

The BREST FIR dataset alone does not provide a sufficient number of flights for large-scale experiments. Thus, we use the full European dataset, grouping sectors by country and treating each as a sector-like agent, as illustrated in \Cref{fig:TDP_Europe} for the scalability test.
\fix{For each traffic volume in $\{10,30,100,300,1000,3000,10000\}$ aircraft, we generate 30 independent instances by uniformly sampling flights at random.} Each agent is assigned a uniform capacity limit $D_i = D$, with $D$ calibrated to represent 85\% of the maximum observed occupancy in each scenario.

\paragraph{Implementation details}
\fix{All experiments were implemented in MATLAB. The local best-response problems and centralized optimization baseline are formulated as mixed-integer linear programs and solved using \texttt{intlinprog}.} Experiments were conducted on a \fix{2024 Mac mini with an Apple M4 Pro chip and 24 GB of memory}.

\subsection{Baselines}
We compare the proposed decentralized algorithm against two representative \fix{centralized} benchmarks.

\begin{enumerate}[\hspace{1pt}1)]
\item \textbf{Centralized solver \fix{(\texttt{CENT})}:}
The centralized solver simulates an idealized system-wide coordinator with authority to modify all flight plans. It solves: 
\begin{equation}\label{eq:centralized-problem}
\begin{aligned}
\min_{\xVec} \quad & \sum_{i\in\plSet} L_i(\xVec) \\
\text{s.t.}\quad & x_i \in \actSet_i,\quad \forall i\in\plSet.
\end{aligned}
\end{equation}
This benchmark represents a centralized reference case under full information sharing and centralized authority.
\fix{We formulate this problem as a mixed-integer linear program and solve it using \texttt{intlinprog}.}

\item \textbf{First-Come-First-Served \fix{(\texttt{FCFS}):}}
\fix{The \texttt{FCFS} heuristic is a centralized sequential baseline based on scheduled departure order. Flights are sequentially inserted into the schedule, and if a flight at its scheduled departure time would introduce an overload on any sector-time pair, the flight is delayed using the same discrete action set $\actSet$ as in our model until an admissible option is found or no further delay options remain. This heuristic captures a simple first-come-first-served scheduling policy without joint optimization over all flights.}
\end{enumerate}

\subsection{Results}
We now present the results of our numerical experiments, building on the setup and baselines described above.

\begin{figure}[hbt!]
    \centering
    \includegraphics[width=0.99\linewidth]{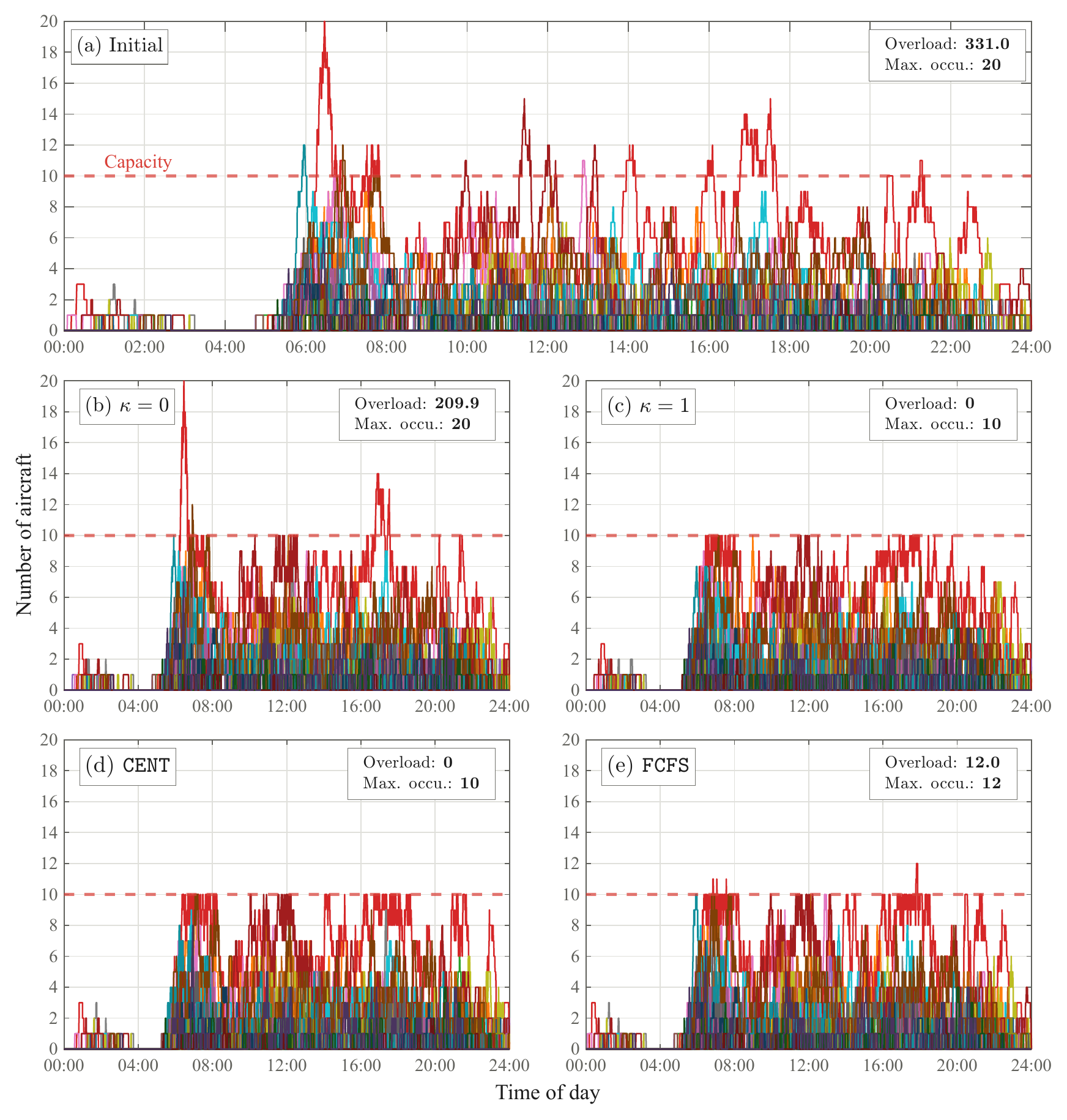}
    \caption{\fix{Traffic demand profiles in the BREST FIR for a representative day with $D=10$. (a) Initial schedule. (b) Resulting schedule for $\kappa=0$. (c) Resulting schedule for $\kappa=1$. (d) Resulting schedule for \texttt{CENT}. (e) Resulting schedule for \texttt{FCFS}.} Side boxes report the total overload [aircraft-min] and maximum instantaneous occupancy [number of aircraft].}
    \label{fig:quals}
\end{figure}

\subsubsection{Qualitative analysis}
We first visually inspect how the proposed algorithm behaves under different $\kappa$ values in a representative case from the BREST FIR dataset \fix{shown in \Cref{fig:quals}}.
\fix{Panels (c) and (d) of \Cref{fig:quals} show that both $\kappa=1$ and \texttt{CENT} eliminate all overload. In contrast, $\kappa=0$ leaves substantial overload unresolved, while \texttt{FCFS} also substantially reduces the initial overload, as shown in panels (b) and (e).}
Having confirmed the algorithm’s behavior in representative cases, we next turn to \fix{the one-month dataset} for a more systematic evaluation, including intermediate values of $\kappa = 10^{-6}$ and $0.5$.

\begin{figure}[hbt!]
    \centering
    \begin{subfigure}{0.49\textwidth}
        \centering
        \includegraphics[width=\linewidth]{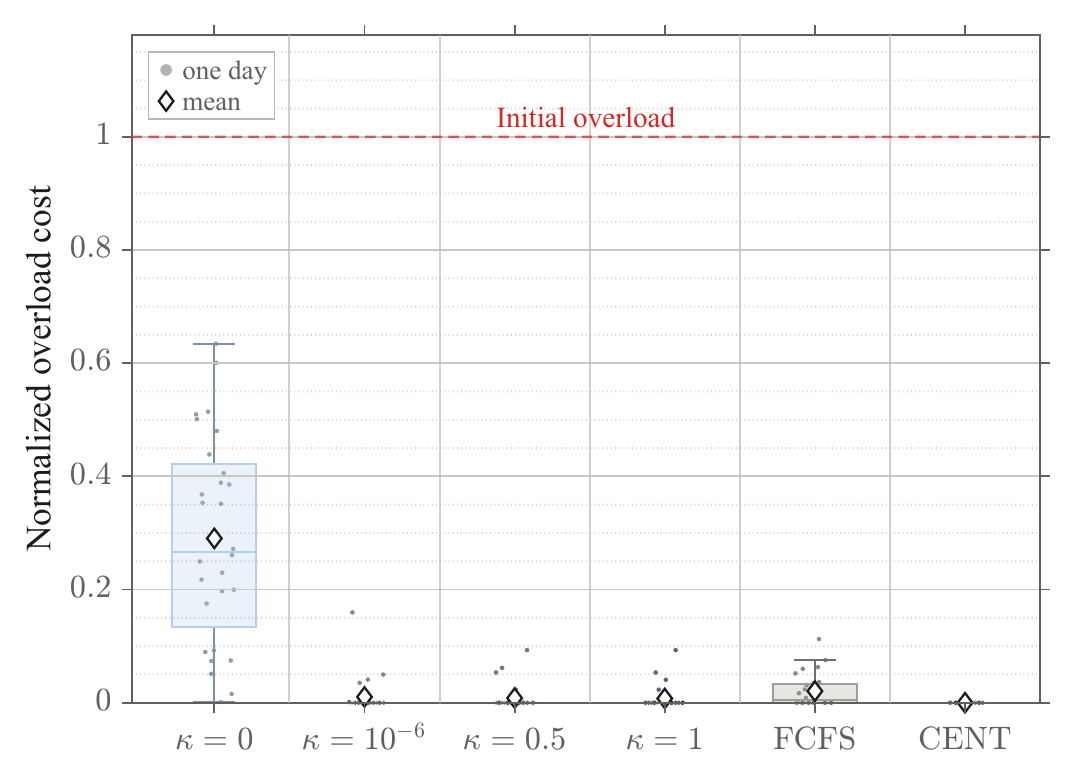}
        \caption{\fix{Normalized overload cost}}
        \label{fig:d10-cost}
    \end{subfigure}
    \hfill
    \begin{subfigure}{0.49\textwidth}
        \centering
        \includegraphics[width=\linewidth]{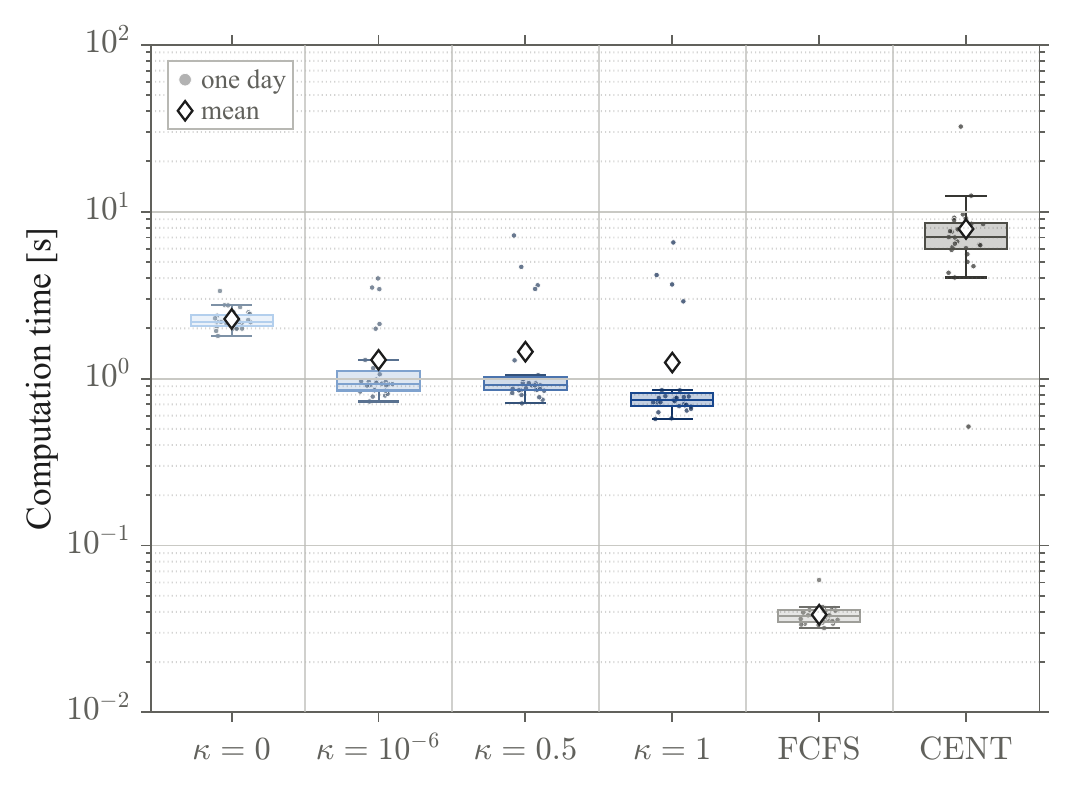}
        \caption{\fix{Computation time}}
        \label{fig:d10-time}
    \end{subfigure}
    \caption{\fix{Performance comparison under the nominal sector capacity $D=10$. (a) Overload cost normalized by the initial overload of each day. (b) Computation time for the proposed method, \texttt{FCFS}, and \texttt{CENT}. Gray points represent individual days, and diamonds indicate the mean.}}
    \label{fig:d10-results}
\end{figure}

\subsubsection{Effect of $\kappa$ on the algorithm performance}

\fix{The decentralized algorithm with $\kappa=0$ reduces the initial overload cost by approximately 71\% on average, whereas the positive $\kappa$ cases achieve approximately 99.0--99.3\% reduction over the one-month dataset, as shown in \Cref{fig:d10-cost}. Even the self-prioritizing cooperation regime ($\kappa=10^{-6}$) achieves performance close to the larger cooperation factors.}

With respect to computation time, \fix{the $\kappa=0$ case takes the longest to converge among the decentralized cases, requiring approximately 1.6--1.8 times the runtime of the positive $\kappa$ cases, as shown in \Cref{fig:d10-time}. Because the self-interested agents do not account for other agents' overload, they often leave residual overload, so the algorithm cannot terminate through the feasibility condition and continues until an additional round confirms that no agent can further improve its cost.}

\subsubsection{Baseline comparison}

Compared with the \fix{centralized baselines} (\fix{\texttt{CENT} and \texttt{FCFS}}), the decentralized algorithm with $\kappa>0$ \fix{achieves lower overload cost than \texttt{FCFS} and approaches the centralized optimization result. The positive $\kappa$ cases reduce the initial overload by approximately 99.0--99.3\%, compared with 98.0\% for \texttt{FCFS} and 100\% for \texttt{CENT}. Thus, the proposed method improves upon \texttt{FCFS} by approximately 1.0--1.3 percentage points in average overload reduction while leaving only 0.7--1.0\% of the initial overload.}

Computation times reveal a \fix{different} pattern. \fix{\texttt{FCFS} is the fastest method, requiring approximately 0.04 s on average, whereas the positive $\kappa$ cases require approximately 1.25--1.45 s. \texttt{CENT} requires approximately 7.9 s on average, or about 5.4--6.3 times longer than the positive $\kappa$ cases, as shown in \Cref{fig:d10-time}. The proposed method therefore achieves overload mitigation close to \texttt{CENT} and slightly better than \texttt{FCFS}, while its computation time lies between the two centralized baselines.}

\begin{figure}[hbt!]
    \centering
    \begin{subfigure}{0.49\textwidth}
        \centering
        \includegraphics[width=\linewidth]{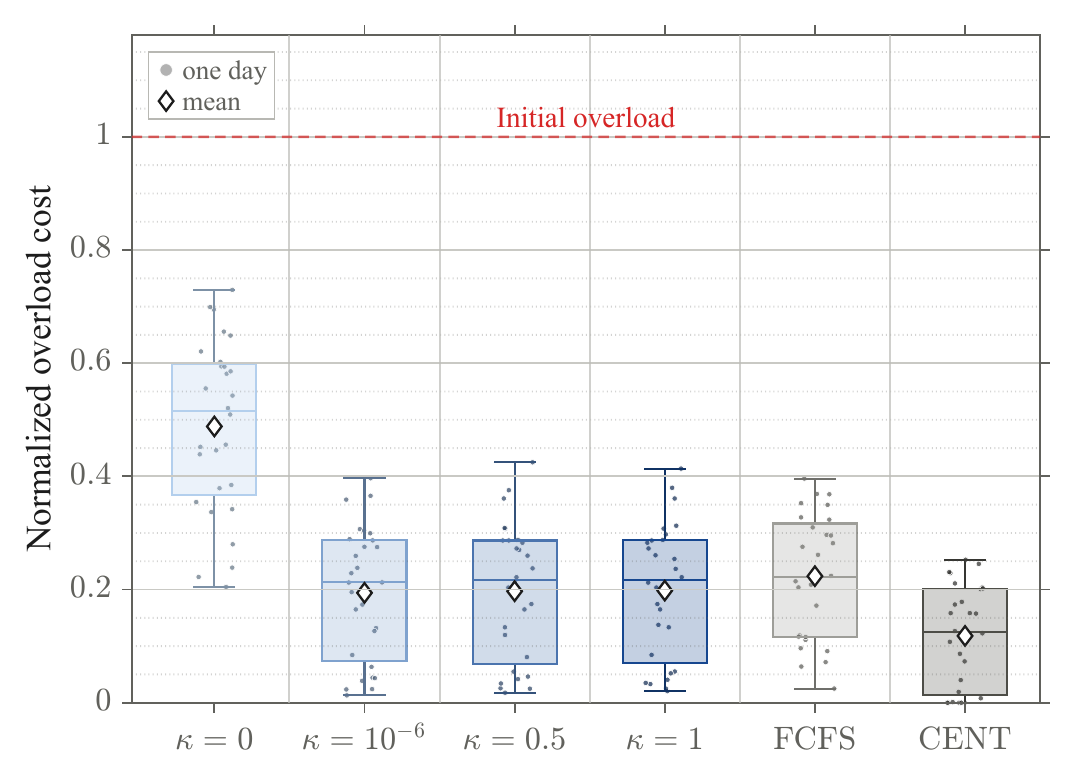}
        \caption{\fix{Normalized overload cost}}
        \label{fig:d7-cost}
    \end{subfigure}
    \hfill
    \begin{subfigure}{0.49\textwidth}
        \centering
        \includegraphics[width=\linewidth]{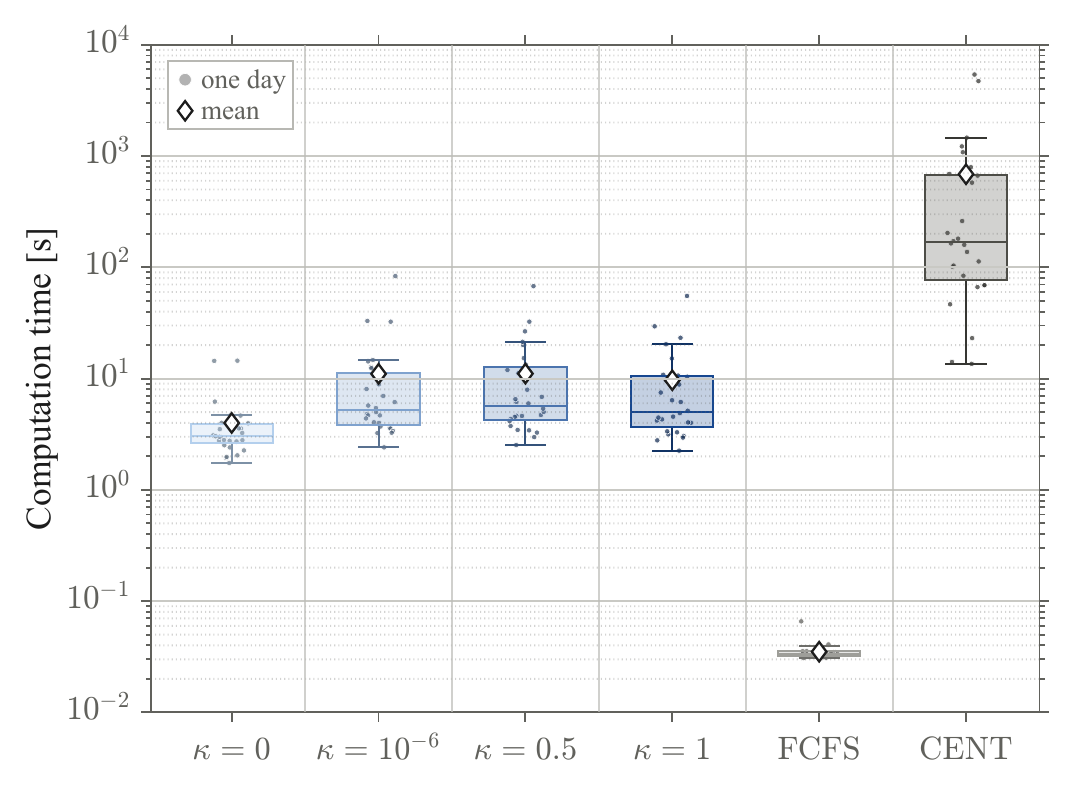}
        \caption{\fix{Computation time}}
        \label{fig:d7-time}
    \end{subfigure}
    \caption{\fix{Performance comparison under the reduced sector capacity $D=7$. (a) Overload cost normalized by the initial overload of each day. (b) Computation time for the proposed method, \texttt{FCFS}, and \texttt{CENT}. Gray points represent individual days, and diamonds indicate the mean.}}
    \label{fig:d7-results}
\end{figure}

\subsubsection{Stress test}

As all $\kappa>0$ cases in the $D=10$ setting \fix{leave only small residual overload}, we additionally evaluate a harder case where finding a feasible solution is difficult. Specifically, we repeat the BREST FIR experiments with a reduced capacity limit of $D=7$, as illustrated in \Cref{fig:d7-results}.

In this setting, \fix{none of the methods consistently identify overload-free solutions. \texttt{CENT} achieves the lowest residual overload, reducing the initial overload cost by approximately 88.2\% on average. The proposed method with $\kappa>0$ achieves approximately 80.2--80.6\% reduction, compared with 77.6\% for \texttt{FCFS}. Thus, the positive $\kappa$ cases improve upon \texttt{FCFS} by approximately 2.6--3.0 percentage points and achieve approximately 91\% of the overload reduction obtained by \texttt{CENT}. The $\kappa=0$ case achieves a substantially smaller reduction of approximately 51.2\%. Among the positive $\kappa$ cases, the differences remain small, with overload-cost reductions differing by only approximately 0.4 percentage points across the tested values.}

The runtime gap between the decentralized and centralized solvers becomes more pronounced in this setting. \fix{\texttt{CENT} requires approximately 62--71 times longer computation on average than the positive $\kappa$ cases, while \texttt{FCFS} remains substantially faster than the proposed method, as shown in \Cref{fig:d7-time}. Thus, under the tighter capacity constraint, the proposed method achieves overload mitigation slightly better than \texttt{FCFS} and close to \texttt{CENT}, while requiring substantially less computation than \texttt{CENT}.}

\begin{figure}[hbt!]
    \centering
    \begin{subfigure}{0.49\textwidth}
        \centering
        \includegraphics[width=\linewidth]{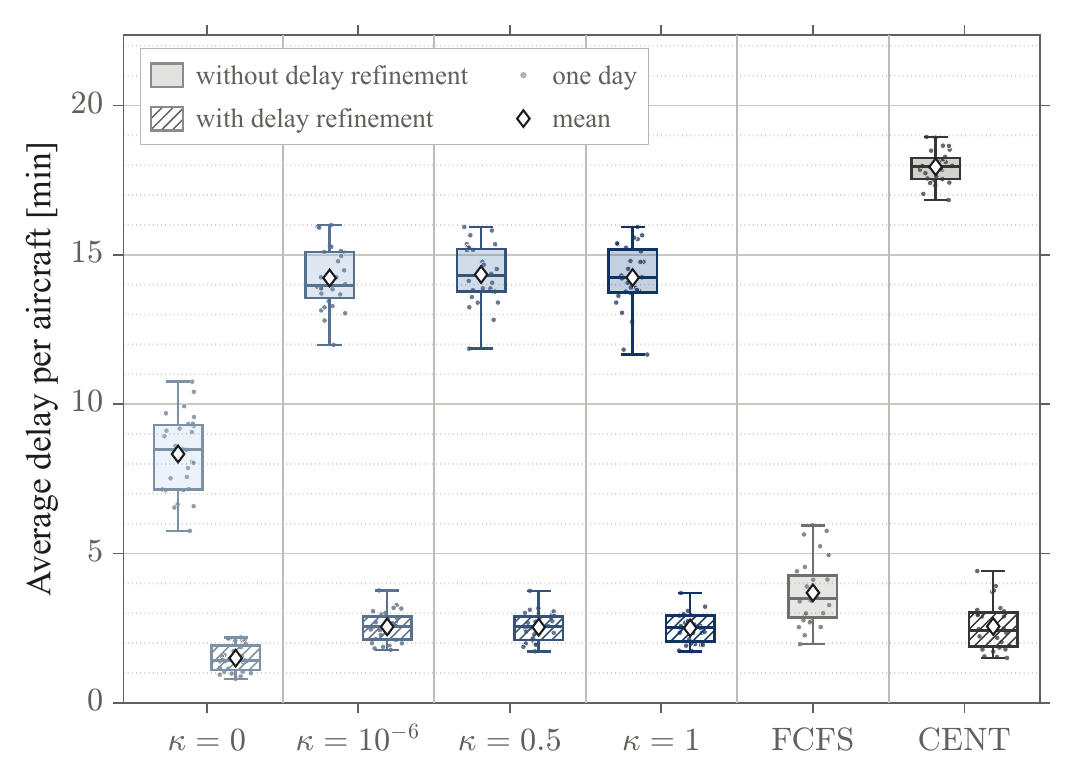}
        \caption{\fix{Average departure delay}}
        \label{fig:delay-analysis}
    \end{subfigure}
    \hfill
    \begin{subfigure}{0.49\textwidth}
        \centering
        \includegraphics[width=\linewidth]{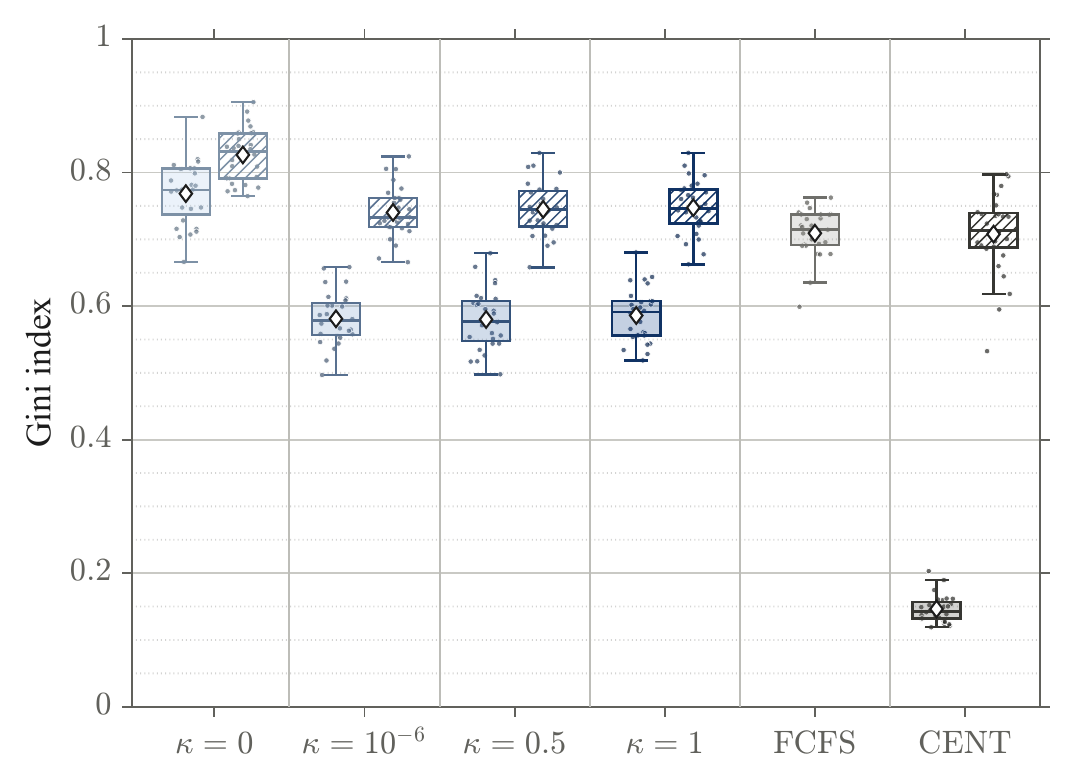}
        \caption{\fix{Gini index of mean delay per sector}}
        \label{fig:gini-analysis}
    \end{subfigure}
    \caption{\fix{Departure delay and fairness comparison. (a) Average departure delay per aircraft. (b) Gini index of mean departure delay across sectors, where a lower Gini index indicates a fairer delay distribution. Solid boxes show the original solutions, while hatched boxes show the solutions after delay refinement. Gray points represent individual days, and diamonds indicate the mean.}}
    \label{fig:delay-fairness}
\end{figure}

\subsubsection{\fix{Delay and fairness analysis}}

\fix{We additionally evaluate departure delay and fairness as operational performance metrics, although the proposed algorithm does not directly optimize either quantity. The positive $\kappa$ cases result in an average departure delay of approximately 14.2 minutes per aircraft (solid boxes in \Cref{fig:delay-analysis}), compared with 3.7 min for \texttt{FCFS} and 17.9 min for \texttt{CENT}. The corresponding total departure delay is approximately 19{,}100--19{,}300 aircraft-min per day for the proposed method.}

\fix{We further apply an additional delay optimization that minimizes departure delay without increasing the overload of any sector. This refinement reduces the average delay of the positive $\kappa$ cases from approximately 14.2--14.3 min to 2.5 min per aircraft, corresponding to an approximately 82\% reduction. The refined solutions achieve lower average delay than \texttt{FCFS} and remain close to the delay-refined \texttt{CENT} solution at 2.6 min, as shown in \Cref{fig:delay-analysis}.}
\fix{The corresponding per-flight delay distributions for all methods are reported in \ref{app:delay-dist}.}

\fix{A lower Gini index indicates a fairer distribution of departure delay across sectors, while a higher value indicates greater inequality.}
\fix{The sector-level Gini index of departure delay increases from approximately 0.58--0.59 to 0.74--0.75 after delay refinement. These values remain comparable to \texttt{FCFS} and the delay-refined \texttt{CENT} solution (both approximately 0.71), as shown in \Cref{fig:gini-analysis}. This indicates a trade-off between reducing aggregate delay and distributing delay evenly across sectors. This analysis shows that substantial delay reduction may be achievable while preserving the same overload level. However, the proposed algorithm does not guarantee convergence to such solutions or that the resulting solution remains a pure Nash equilibrium. Developing decentralized dynamics that naturally achieve such outcomes is an interesting direction for future work.}

\begin{figure}[hbt!]
    \centering
    \includegraphics[width=0.6\linewidth]{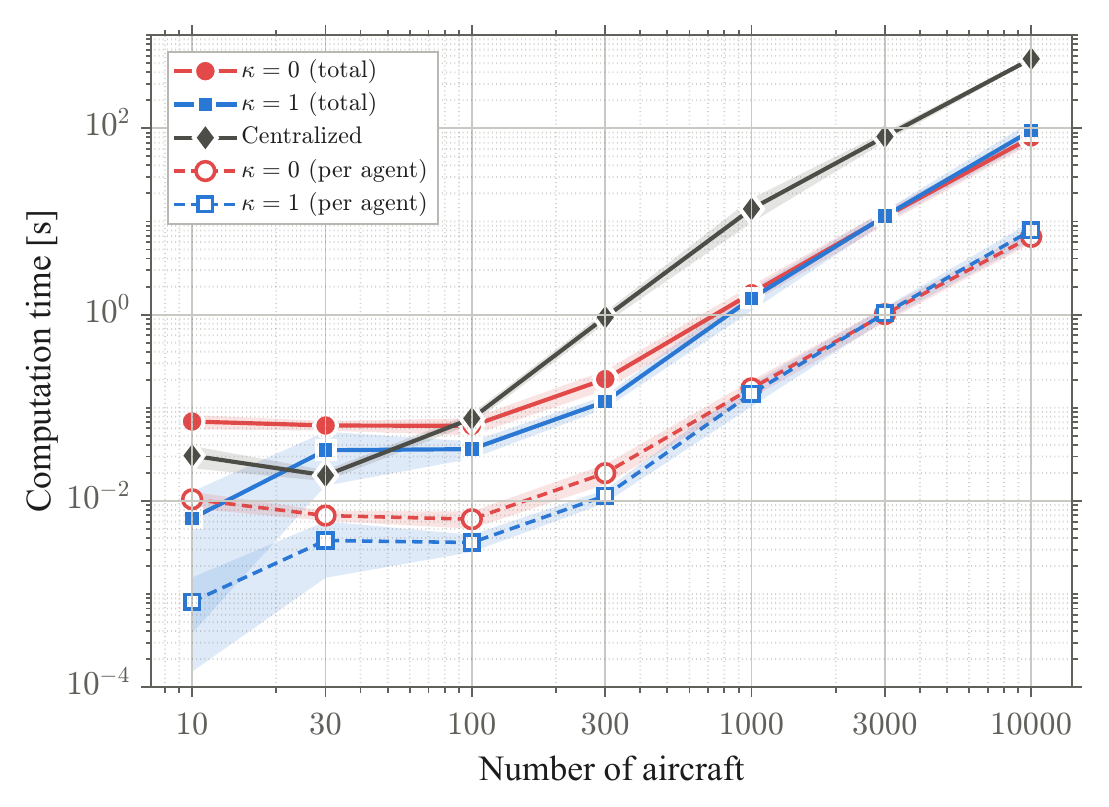}
    \caption{\fix{Scalability comparison between the proposed method and \texttt{CENT}. Solid lines show the mean total computation time, while dashed lines show the mean per-agent computation time for the proposed method. Shaded regions indicate 95\% confidence intervals for the mean over 30 independently sampled instances at each traffic volume.}}
    \label{fig:scalability}
\end{figure}

\subsubsection{Scalability}

We evaluate the scalability of the proposed algorithm with varying numbers of aircraft.
To test scenarios involving larger traffic volumes, we utilize the full European dataset aggregated into 12 country-level agents with capacity limits set to 85\% of the maximum observed occupancy, as described in \Cref{sec:exp-setup}.
\fix{We sample 30 random instances for each traffic volume ranging from 10 to 10,000 flights.}

As shown in \Cref{fig:scalability}, \fix{the total runtimes of the decentralized algorithm grow at a rate similar to that of \texttt{CENT} but remain substantially lower in the larger instances. At 10,000 aircraft, \texttt{CENT} requires approximately 622 s, compared with approximately 64 s for the proposed method with $\kappa=1$, corresponding to nearly one order of magnitude lower computation time.}
The distribution of computation in the decentralized algorithm leads to substantially lower per-agent runtimes (dashed lines), \fix{which remain up to two orders of magnitude below the \texttt{CENT} runtime}.
This highlights the reduced computational burden per participant in the decentralized approach.

\section{Discussion}

\subsection{Minimal cooperation as a key driver \fix{of regulated decentralization}}
Experiments show that the decentralized algorithm achieves effective overload mitigation with only a minimal degree of cooperation. 
The fully self-interested case ($\kappa=0$) consistently \fix{performs worse}, whereas the self-prioritizing regime ($\kappa=10^{-6}$) already \fix{substantially reduces overload, as shown in \Cref{fig:d10-cost} and \Cref{fig:d7-cost}}. 
Thus, full altruism is unnecessary; a minimal willingness to act beyond one’s own benefit \fix{can substantially improve overload mitigation}.

In terms of runtime, \fix{the decentralized algorithm requires substantially less computation than the centralized solver (\texttt{CENT}) in the tested settings, while \texttt{FCFS} remains computationally faster.} 

\fix{Overall, the proposed algorithm occupies a useful intermediate regime. It achieves solution quality close to centralized coordination while requiring substantially less computation than \texttt{CENT} and preserving distributed decision authority. 
This distinction is important in settings where stakeholders have different or conflicting objectives and centralized schedule authority may be difficult to establish. 
Despite agents acting through local best responses, the results show that performance close to centralized coordination can be achieved with (i) a minimal degree of cooperation, namely the self-prioritizing cooperation characterized in \Cref{thm:self-prio}, while (ii) the common no-new-overload rule supports convergence of the regulated best-response dynamics.}

\subsection{Empirical performance versus theoretical guarantees}
\Cref{thm:feasible} guarantees that when $\kappa=1$, the minimizer of the potential function corresponds to a feasible solution if such a solution exists. However, this does not ensure that best-response dynamics will actually reach such \fix{a} solution. 
In practice, our experiments show that feasible solutions were often obtained even for $\kappa \in (0,1)$, outside the scope of the theoretical guarantee. This observation highlights the practical effectiveness of the algorithm and motivates future work on closing the gap between theoretical guarantees and observed performance.

\subsection{Implications for real-world deployment}

Using real traffic data, we demonstrated that the proposed regulated decentralized algorithm effectively reduces overload. The setting assumes shared flight-plan and surveillance data so that sector-time occupancies, overloads, and admissibility of candidate updates can be evaluated consistently. 
This is a sensible operational assumption, as \fix{surveillance information can be made available} through systems such as automatic dependent surveillance--broadcast (ADS-B)\fix{, together with shared flight-plan information}.
Decentralization in this framework therefore refers to the decision-making structure, not to the absence of shared traffic information.

The agents in this framework can represent sectors, sector groups, flight information regions (FIRs), or functional airspace blocks (FABs), depending on the level at which coordination is implemented. 
The no-new-overload restriction should be interpreted as a sufficient protocol-level regulation for stable operation. It restricts local best-response updates so that convergence to a \fix{pure-strategy Nash} equilibrium can be guaranteed, providing the stability needed for \fix{the distributed coordination mechanism considered here}.

The main operational implication is that effective overload mitigation does not require full cooperation or strong altruism among agents. The experiments indicate that even a very small willingness to account for other agents' overload, as long as the agent does not worsen its own overload, can substantially improve system performance. This suggests that regulated decentralized ATM can be practically meaningful in settings where agents retain local decision authority but follow a common no-new-overload protocol.

The present formulation is still an abstraction of real ATM operations. We use fixed sector capacities, an occupancy-based overload metric, and departure-time adjustments as the control action. 
\fix{In particular, sector occupancy should not be interpreted as a direct measure of controller cognitive workload. A schedule that avoids occupancy overload may still maintain traffic near nominal capacity for sustained periods and therefore impose increased workload. We provide in \ref{app:workload} a post-hoc evaluation using simplified dynamic density as a workload proxy.
More generally, once an appropriate workload metric is specified, the proposed framework can accommodate an additional constraint on candidate schedule updates to limit workload.} 
Future extensions should incorporate \fix{workload or traffic-complexity measures,} dynamic sector configurations such as sector opening and collapsing, time-varying capacities, rerouting decisions, and richer complexity metrics beyond occupancy.

\section{Conclusion}
We propose a game-theoretic regulated decentralized protocol for mitigating sector overload in air traffic management (ATM). The model introduces a tunable cooperativeness factor $\kappa$, which interpolates between fully self-interested behavior ($\kappa=0$) and fully cooperative behavior ($\kappa=1$). 
We show that the resulting game admits an unconditional potential structure when $\kappa=1$, and that, for $0\leq\kappa<1$, restricted best-response dynamics terminates in finite time under a no-new-overload restriction. Leveraging these properties, we prove that best-response dynamics converges to a pure Nash equilibrium under the stated conditions.

Numerical experiments with real European flight data confirm the practical effectiveness of the approach. The algorithm \fix{substantially reduces overload for positive values of $\kappa$, including very small values corresponding to self-prioritizing cooperation that does not worsen an agent's own overload.}
Stress tests with tighter capacities and scalability tests with up to 10,000 flights further show that the decentralized algorithm \fix{achieves overload mitigation close to centralized optimization while requiring considerably less computation than \texttt{CENT}. These results show that distributed decision authority can retain solution quality comparable to centralized coordination when combined with minimal cooperation and the no-new-overload rule.}

These results highlight the central insight that effective decentralized coordination does not require full altruism. Minimal cooperation that does not compromise one’s own interest is sufficient to \fix{substantially improve system-level overload mitigation and approach the performance of centralized coordination.} Future work will further strengthen theoretical guarantees for convergence to overload-free solutions. Future research directions include integrating the framework with rerouting strategies and extending it to account for various measures of sector complexity beyond the occupancy metric \cite{occupancy_count_1, occupancy_count_2}, as well as incorporating uncertainties in demand and capacity forecasts.

\section*{CRediT authorship contribution statement}
\noindent
\textbf{Jaehan Im}: Conceptualization, Methodology, Software, Formal analysis, Writing – original draft. 
\textbf{Daniel Delahaye}: Supervision, Resources, Validation, Writing – review \& editing. 
\textbf{David Fridovich-Keil}: Supervision, Funding acquisition, Methodology, Writing – review \& editing. 
\textbf{Ufuk Topcu}: Supervision, Funding acquisition, Methodology, Writing – review \& editing.

\section*{Declaration of competing interest}
The authors declare that they have no known competing financial interests or personal relationships that could have appeared to influence the work reported in this paper.

\section*{Acknowledgments}
This work was supported by the National Science Foundation CAREER award under grants 2336840 and 2211548, by the National Aeronautics and Space Administration ULI Award under grants 80NSSC21M0071 and 80NSSC24M0070, and by the Office of Naval Research under grant N00014-22-1-2703.

\section*{Data availability}
The data that has been used is confidential.

\appendix
\section{Proofs for the theorems and lemmas} \label{app1}

\begin{proof}[\textbf{Proof of \Cref{thm:self-prio}}] \label{app:self-prio}
Recall the cost function from \Cref{eq:cost}:
\begin{equation}
    J_i(\xVec) = L_i(\xVec) + \kappa \sum_{j\in\plSet\setminus\{i\}} L_j(\xVec).
\end{equation}
For a unilateral deviation $x_i\to x_i'$, define
\begin{equation}
   \Delta L_i = L_i(\xVec_i') - L_i(\xVec), 
   \quad 
   \Delta L_{- i} = \sum_{j\ne i}\!\big(L_j(\xVec_i') - L_j(\xVec)\big).
\end{equation}
Then
\begin{equation}
   J_i(\xVec_i') - J_i(\xVec)
   = \Delta L_i + \kappa \Delta L_{- i}.
\end{equation}

Consider the critical case in which \fix{agent} $i$'s own cumulative overload increases, i.e., $\Delta L_i>0$. Since overload is evaluated over time bins of length $\Delta\tau$, any positive increase in cumulative overload is at least $\Delta\tau$. Hence,
\begin{equation}
    \Delta L_i \geq \Delta\tau.
\end{equation}
\fix{On the other hand, a unilateral deviation by agent $i$ changes only the occupancy contributions of flights in $\flSet_i$. For each sector $j\neq i$, moving such a flight can reduce cumulative overload by at most the total time that the flight spends in that sector. Since departure-time adjustments shift the sector entry and exit times by the same amount, these dwell durations are preserved. Summing over all flights assigned to agent $i$ and all other sectors therefore gives
\begin{equation}
    \Delta L_{-i} \geq -W_i,
\end{equation}
where $W_i$ is defined in \Cref{eq:Wi}.}
Therefore, whenever $\Delta L_i>0$,
\begin{equation}
   J_i(\xVec_i') - J_i(\xVec)
   \geq
   \fix{\Delta\tau - \kappa W_i}.
\end{equation}
If
\begin{equation}
   \fix{\kappa < \frac{\Delta\tau}{W_i}},
\end{equation}
then $J_i(\xVec_i')-J_i(\xVec)>0$ for any deviation that increases \fix{agent} $i$'s own cumulative overload. Such a deviation cannot be cost-improving and therefore will not be adopted. \fix{Requiring this condition for every agent gives
\begin{equation}
   \kappa < \frac{\Delta\tau}{\max_{i\in\plSet}W_i}.
\end{equation}
Hence, no agent adopts an action that increases its own cumulative overload, completing the proof.}
\end{proof}

\begin{proof}[\textbf{Proof of \Cref{thm:potential-fixed}}] \label{app:potential-fixed}
Recall the cost function in \Cref{eq:cost}:
\begin{equation}
    J_i(\xVec) = L_i(\xVec) + \kappa \sum_{j \neq i} L_j(\xVec).
\end{equation}
Consider a unilateral deviation $x_i\to x_i'$ and suppose that the overloaded sector-time set remains unchanged (\ie, $\mathcal{O}(\xVec_i')=\mathcal{O}(\xVec)$).
For compactness, define
\begin{equation}
    \Delta C_{ij}^{\tau}
    =
    C_{ij}^{\tau}(\xVec_i') - C_{ij}^{\tau}(\xVec).
\end{equation}
By \Cref{eq:Cjk_invariance}, only the occupancy contributions generated by flights scheduled by sector $i$ can change. Hence, for any sector $j$,
\begin{equation}
    L_j(\xVec_i')-L_j(\xVec)
    =
    \sum_{\tau:(j,\tau)\in\mathcal{O}(\xVec)}
    \Delta C_{ij}^{\tau}\Delta\tau,
\end{equation}
because the active overloaded sector-time set is fixed.

The cost difference for player $i$ is therefore
\begin{equation}
\begin{aligned}
    J_i(\xVec_i')-J_i(\xVec)
    &=
    L_i(\xVec_i')-L_i(\xVec)
    +\kappa\sum_{j\in\plSet\setminus\{i\}}\left(L_j(\xVec_i')-L_j(\xVec)\right)\\
    &=
    \sum_{\tau:(i,\tau)\in\mathcal{O}(\xVec)}
    \Delta C_{ii}^{\tau}\Delta\tau
    +
    \kappa
    \sum_{j\in\plSet\setminus\{i\}}
    \sum_{\tau:(j,\tau)\in\mathcal{O}(\xVec)}
    \Delta C_{ij}^{\tau}\Delta\tau .
\end{aligned}
\end{equation}

Now consider the potential function in \Cref{eq:potential}:
\begin{equation}
    \Phi(\xVec)
    =
    \kappa\sum_{j\in\plSet}L_j(\xVec)
    +
    (1-\kappa)
    \sum_{(j,\tau)\in\mathcal{O}(\xVec)}
    C_{jj}^{\tau}(\xVec)\Delta\tau .
\end{equation}
Since $\mathcal{O}(\xVec_i')=\mathcal{O}(\xVec)$, its difference under the unilateral deviation is
\begin{equation}
\begin{aligned}
    \Phi(\xVec_i')-\Phi(\xVec)
    &=
    \kappa\sum_{j\in\plSet}\left(L_j(\xVec_i')-L_j(\xVec)\right)\\
    &\quad+
    (1-\kappa)
    \sum_{(j,\tau)\in\mathcal{O}(\xVec)}
    \left(C_{jj}^{\tau}(\xVec_i')-C_{jj}^{\tau}(\xVec)\right)\Delta\tau .
\end{aligned}
\end{equation}
The second term changes only for $j=i$, because sector $i$'s update does not alter the flights scheduled by other sectors. Thus,
\begin{equation}
\begin{aligned}
    \Phi(\xVec_i')-\Phi(\xVec)
    &=
    \kappa
    \sum_{j\in\plSet}
    \sum_{\tau:(j,\tau)\in\mathcal{O}(\xVec)}
    \Delta C_{ij}^{\tau}\Delta\tau\\
    &\quad+
    (1-\kappa)
    \sum_{\tau:(i,\tau)\in\mathcal{O}(\xVec)}
    \Delta C_{ii}^{\tau}\Delta\tau\\
    &=
    \sum_{\tau:(i,\tau)\in\mathcal{O}(\xVec)}
    \Delta C_{ii}^{\tau}\Delta\tau
    +
    \kappa
    \sum_{j\in\plSet\setminus\{i\}}
    \sum_{\tau:(j,\tau)\in\mathcal{O}(\xVec)}
    \Delta C_{ij}^{\tau}\Delta\tau .
\end{aligned}
\end{equation}
This equals $J_i(\xVec_i')-J_i(\xVec)$. Therefore, $\Phi$ is a potential function whenever the overloaded sector-time set remains fixed.
\end{proof}

\begin{proof}[\textbf{Proof of \Cref{lem:kappa-one}}] \label{app:kappa-extremes}
When $\kappa=1$, each player's cost becomes
\begin{equation}
    J_i(\xVec) = \sum_{j \in \plSet} L_j(\xVec), \quad \forall i.
\end{equation}
Define
\begin{equation}
    \Phi(\xVec) = \sum_{j \in \plSet} L_j(\xVec).
\end{equation}
Then, for any unilateral deviation $x_i\to x_i'$,
\begin{equation}
    J_i(\xVec_i') - J_i(\xVec)
    =
    \Phi(\xVec_i') - \Phi(\xVec).
\end{equation}
Therefore, the game admits $\Phi$ as an unconditional potential function when $\kappa=1$.
\end{proof}

\begin{proof}[\textbf{Proof of \Cref{lem:monotone}}]\label{app:monotone}
Under the no-new-overload restriction in \Cref{def:noNewOvld},
\begin{equation}
    \mathcal{O}_f(\xVec,\xVec_i')=\emptyset.
\end{equation}
By definition of $\mathcal{O}_f$, this means that no sector-time pair that was feasible before the deviation becomes overloaded after the deviation. Equivalently,
\begin{equation}
    \mathcal{O}(\xVec_i')\setminus \mathcal{O}(\xVec)=\emptyset,
\end{equation}
and hence
\begin{equation}
    \mathcal{O}(\xVec_i')\subseteq \mathcal{O}(\xVec).
\end{equation}
Therefore, if the overloaded sector-time set changes, i.e.,
\begin{equation}
    \mathcal{O}(\xVec_i')\neq \mathcal{O}(\xVec),
\end{equation}
then the inclusion must be strict:
\begin{equation}
    \mathcal{O}(\xVec_i')\subsetneq \mathcal{O}(\xVec).
\end{equation}
This proves the claim.
\end{proof}

\begin{proof}[\textbf{Proof of \Cref{thm:termination}}]\label{app:termination}
Consider the restricted best-response dynamics under the no-new-overload restriction. By \Cref{lem:monotone}, the overloaded sector-time set satisfies
\begin{equation}
    \mathcal{O}(\xVec^{(t+1)})\subseteq \mathcal{O}(\xVec^{(t)})
\end{equation}
for every accepted unilateral update. Moreover, if the overloaded sector-time set changes, then
\begin{equation}
    \mathcal{O}(\xVec^{(t+1)})\subsetneq \mathcal{O}(\xVec^{(t)}).
\end{equation}
Since $\plSet\times\mathcal{T}$ is finite, such strict decreases can occur only finitely many times.

It remains to consider iterations during which $\mathcal{O}(\xVec^{(t)})$ remains fixed. For any fixed overloaded sector-time set, \Cref{thm:potential-fixed} shows that the restricted game admits a potential function. Since the action space is finite and accepted updates are cost-improving, the potential decreases along the update sequence and the process cannot continue indefinitely while $\mathcal{O}(\xVec^{(t)})$ is fixed.

Combining the two observations, the restricted best-response dynamics must terminate in finite time. At termination, no admissible unilateral update can further reduce any agent's cost. Therefore, the terminal joint action is a pure Nash equilibrium with respect to the admissible action set induced by \Cref{def:noNewOvld}.
\end{proof}

\begin{proof}[\textbf{Proof of \Cref{thm:convergence}}] \label{app:convergence}
When $\kappa=1$, \Cref{lem:kappa-one} shows that the game admits an unconditional potential function. Since the action space is finite, best-response dynamics with cost-improving updates converges in finite time to a pure Nash equilibrium.

When $0\leq\kappa<1$, \Cref{thm:termination} shows that, under the no-new-overload restriction, the restricted best-response dynamics terminates in finite time. At termination, no admissible unilateral cost-improving update remains. Hence, the terminal joint action is a pure Nash equilibrium with respect to the admissible action set induced by \Cref{def:noNewOvld}.

Therefore, best-response dynamics converges in finite time for $\kappa=1$ unconditionally and for $0\leq\kappa<1$ under the no-new-overload restriction.
\end{proof}

\begin{proof}[\textbf{Proof of \Cref{thm:feasible}}] \label{app:feasible}
For $\kappa=1$, the cost function simplifies to
\begin{equation}
    J_i(\xVec) = \sum_{j \in \plSet} L_j(\xVec), \quad \forall i,
\end{equation}
and the potential function is the same:
\begin{equation}
    \Phi(\xVec) = \sum_{j\in\plSet} L_j(\xVec).
\end{equation}

If $\xVec^\star$ is feasible, then $L_j(\xVec^\star)=0$ for all $j$, hence 
\begin{equation}
    \Phi(\xVec^\star) = 0.
\end{equation}
Because overloads are nonnegative, $0$ is the minimum possible value of $\Phi$.  
Therefore any feasible solution is a global minimizer of the potential.
\end{proof}



\section{\fix{Per-flight departure delay distribution}} \label{app:delay-dist}

\fix{The delay refinement substantially shifts the distribution toward smaller departure delays. For the positive-$\kappa$ cases, the average share of flights receiving no departure delay increases from approximately 38\% before refinement to approximately 84\% after refinement. The refined distributions are also similar across the positive-$\kappa$ values, consistent with the average-delay results reported in the main text.}

\begin{table}[hbt!]
\centering
\caption{\fix{Per-flight departure delay distribution under the reduced sector capacity $D=7$. 
Entries report the share of flights [\%] assigned to each departure-delay value, averaged over the 28 experimental days. 
``w/o'' and ``w/'' denote solutions before and after delay refinement, respectively.}}
\label{tab:delay-dist}

\setlength{\tabcolsep}{4pt}
\begin{tabular}{llrrrrrrr|r}
\toprule
 & & \multicolumn{7}{c|}{Departure delay [min]} & Mean \\
Method & & 0 & 5 & 10 & 15 & 20 & 25 & 30 & [min] \\
\midrule
$\kappa=0$       & w/o & 63.4 & 3.1 & 3.8 & 3.6 & 3.7 & 4.3 & 18.1 & 8.3 \\
                 & w/  & 89.9 & 3.0 & 2.0 & 1.3 & 1.2 & 1.1 & 1.4  & 1.5 \\
$\kappa=10^{-6}$ & w/o & 38.3 & 5.2 & 6.0 & 5.6 & 5.9 & 7.0 & 31.9 & 14.2 \\
                 & w/  & 83.6 & 4.7 & 3.2 & 2.1 & 1.8 & 1.7 & 2.9  & 2.5 \\
$\kappa=0.5$     & w/o & 37.8 & 5.3 & 6.0 & 5.7 & 6.0 & 7.1 & 32.2 & 14.3 \\
                 & w/  & 83.5 & 4.8 & 3.2 & 2.1 & 1.8 & 1.8 & 2.8  & 2.5 \\
$\kappa=1$       & w/o & 38.1 & 5.2 & 6.1 & 5.8 & 5.8 & 6.9 & 32.0 & 14.2 \\
                 & w/  & 83.6 & 4.8 & 3.1 & 2.1 & 1.8 & 1.8 & 2.8  & 2.5 \\
\texttt{CENT}    & w/o & 16.5 & 9.0 & 9.5 & 9.4 & 10.0 & 10.9 & 34.7 & 17.9 \\
                 & w/  & 83.5 & 5.3 & 2.8 & 1.8 & 1.5 & 1.7 & 3.4  & 2.6 \\
\texttt{FCFS} & & 77.5 & 5.3 & 4.2 & 3.3 & 2.8 & 2.3 & 4.5  & 3.7 \\
\bottomrule
\end{tabular}
\end{table}

\section{\fix{Post-hoc workload evaluation}} \label{app:workload}
\begin{table}[hbt!]
\centering
\caption{\fix{Post-hoc comparison of daily simplified dynamic density (SDD). Values in parentheses indicate the relative change from the initial schedule.}}
\label{tab:sdd}
\fix{
\begin{tabular}{lccccc}
\hline
 & Initial & $\kappa=0$ & $\kappa=1$ & \texttt{CENT} & \texttt{FCFS} \\
\hline
Mean
& 26043
& 26055 $(+0.05\%)$
& 26077 $(+0.13\%)$
& 26105 $(+0.24\%)$
& 26032 $(-0.04\%)$ \\
Median
& 25484
& 25375 $(-0.43\%)$
& 25422 $(-0.24\%)$
& 25557 $(+0.29\%)$
& 25460 $(-0.10\%)$ \\
Maximum
& 32512
& 32400 $(-0.34\%)$
& 32610 $(+0.30\%)$
& 32792 $(+0.86\%)$
& 32492 $(-0.06\%)$ \\
\hline
\end{tabular}
}
\end{table}

\fix{To assess whether occupancy-overload mitigation substantially changes traffic complexity, we additionally evaluate the resulting schedules using the simplified dynamic density (SDD) metric proposed in \cite{occupancy_count_2}.
Table~\ref{tab:sdd} reports the mean, median, and maximum daily SDD over the one-month dataset, together with the relative change from the initial schedule.}   

\fix{Remark: The SDD weights in \cite{occupancy_count_2} were calibrated using U.S. airspace and are not specific to the BREST FIR. We therefore use SDD only as a relative workload proxy.
In addition, the reference does not fully specify the implementation of several components. 
For the traffic-density component, we use the sector volume in NM$^{3}$ and average the resulting aircraft density over each 15-min period. 
For the heading- and speed-variance components we pool the one-minute
observations of a 15-min period and compute a single variance per sector and
period.}

\bibliographystyle{elsarticle-num-names}
\bibliography{reference}

\end{document}